\documentclass{aastex631}

\newcommand{\virga}{\texttt{virga}}  
  
\newcommand{\fsed}{f_{\text{sed}}}
\newcommand{\kzz}{K_{zz}} 
\newcommand{\vf}{v_f} 
\newcommand{\w}{w_*} 
\newcommand{\reff}{r_\text{eff}} 
\newcommand{\zT}{z_{\text{T}}} 
\newcommand{\zB}{z_{\text{B}}} 
\newcommand{\qvs}{q_{vs}} 

\usepackage{mathtools}
\usepackage{amsmath, amsthm, amssymb, amsfonts}
\usepackage{graphicx}
\usepackage[caption=false]{subfig}

\shorttitle{A New Sedimentation Model for Greater Cloud Diversity}
\shortauthors{Rooney et al.}
\graphicspath{{./}{figures/}}
\begin{document}

\title{A New Sedimentation Model for Greater Cloud Diversity in Giant Exoplanets and Brown Dwarfs}

\correspondingauthor{Caoimhe M. Rooney}
\email{caoimhe.m.rooney@nasa.gov}

\author[0000-0001-9005-2872]{Caoimhe M. Rooney}
\affiliation{NASA Ames Research Center,
Moffett Field,
CA, 94035 USA}

\author{Natasha E. Batalha}
\affiliation{NASA Ames Research Center,
Moffett Field, 
CA, 94035 USA}

\author{Peter Gao}
\affiliation{Earth and Planets Laboratory, Carnegie Institution for Science, 5241 Broad Branch Road, NW, Washington, DC 20015, USA}
\affiliation{Department of Astronomy and Astrophysics, University of California Santa Cruz, 1156 High St, Santa Cruz, CA 95064, USA }
\affiliation{NHFP Sagan Fellow}

\author{Mark S. Marley}
\affiliation{Department of Planetary Sciences, Lunar and Planetary Laboratory, University of Arizona, Tucson AZ 85721}

\begin{abstract}

The observed atmospheric spectrum of exoplanets and brown dwarfs depends critically on the presence and distribution of atmospheric condensates. The \cite{ackerman2001cloud} methodology for predicting the vertical distribution of condensate particles is widely used to study cloudy atmospheres and has recently been implemented in an open-source python package $\virga$. The model relies upon input parameter $\fsed$, the sedimentation efficiency, which until now has been held constant. The relative simplicity of this model renders it useful for retrieval studies due to its rapidly attainable solutions. However, comparisons with more complex microphysical models such as CARMA have highlighted inconsistencies between the two approaches, namely that the cloud parameters needed for radiative transfer produced by $\virga$ are dissimilar to those produced by CARMA. To address these discrepancies, we have extended the original Ackerman and Marley methodology in $\virga$ to allow for non-constant $\fsed$ values, in particular those that vary with altitude. We discuss one such parameterization and compare the cloud mass mixing ratio produced by $\virga$ with constant and variable $\fsed$ profiles to that produced by CARMA. We find that the variable $\fsed$ formulation better captures the profile produced by CARMA with heterogeneous nucleation, yet performs comparatively to constant $\fsed$ for homogeneous nucleation. In general, $\virga$ has the capacity to handle any $\fsed$ with an explicit anti-derivative, permitting a plethora of alternative cloud profiles that are otherwise unattainable by constant $\fsed$ values. The ensuing flexibility has the potential to better agree with increasingly complex models and observed data. 

\end{abstract}

\section{Introduction and motivation}
Accurate predictions of the composition of exoplanet and brown dwarf atmospheres is critical to understanding their formation and evolution.
However, a crucial difficulty in robustly determining atmospheric properties from transmission, emission, and reflected light spectroscopy is the presence and distribution of clouds and hazes.

In transmission, clouds mask the gaseous content of the atmosphere, resulting in varying degrees of spectral feature muting \citep{kreidberg2014clouds, sing2016continuum}, which may probe the formation of clouds as a function of irradiation temperature \citep{gao2020aerosol}. Additionally, there is a well-known degeneracy between atmospheric metallicty and cloud coverage that impedes the accurate and precise constraints of atmospheric abundances \citep{line2016influence}. 

In thermal emission, the emergent flux from optically thin window regions emanates from the hot high pressure area. Clouds make those window regions opaque such that the emergent flux emanates from higher altitude, usually cooler layers
\citep{marley2002clouds}. Understanding the formation of clouds as a function of temperature and gravity has been a major focus of several brown dwarf, \citep[e.g.][]{marley2002clouds, leggett2010mid, apai2013hst,allard2012models, morley2012neglected,faherty2014signatures, helling2014atmospheres,Burningam2021cloud}, hot Jupiter \citep[e.g.][]{lee2017dynamic, Parmentier2021, gao2021universal}, and young giant planet \citep[e.g.][]{burrows2004spectra, helling2014disk, barman2011clouds}  observations and theoretical investigations. 

In reflected light, clouds and hazes dictate the zeroth order shape of the spectrum \citep{Batalha2018}.
A thick, highly scattering cloud, such as a water cloud, will produce a high albedo spectrum (a$>$0.7) throughout the near-infrared ($<1 \mu$m), as opposed to a cloud-free spectrum which darkens toward 1$\mu$m \citep{marley2013clouds}. 
This was noted by \cite{marley1999reflected} and \cite{sudarsky2000albedo} and further investigated by \cite{cahoy2010exoplanet}. On the other hand, hazes suppress blue light ($<0.5 \mu$m) otherwise scattered due to Rayleigh. This effect is seen in Jupiter \citep{karkoschka1994spectrophotometry}, and has been studied with regards to exoplanets \citep{gao2017sulfur}. Most recently, \citet{mukherjee2021cloud} showed that our ability retrieve atmospheric properties from observations of exoplanets in reflected light with future flagship missions will strongly depend on our understanding of clouds and hazes.

Overall, it is clear that our understanding of clouds will directly impact many facets of planet and brown dwarf spectroscopy. 
Fully capturing cloud formation behaviour requires an understanding of the microphysical processes governing the formation and evolution of clouds; namely, homogeneous and heterogeneous nucleation, condensation, evaporation, coagulation and transport.
These mechanisms control the size and spatial distributions of condensate particles from which we can determine the essential parameters for radiative transfer: the single scattering albedo, asymmetry parameter and optical depth.
They are also critically dependent upon the characteristics of the atmosphere itself, such as the temperature, chemical abundance and atmospheric dynamics.
Models that incorporate all of the microphysical mechanisms listed above are highly complex and numerically expensive, therefore, there is a need for parametric models that provide more computationally accessible cloud predictions.

Extensive studies into detailed, microphysical modeling have been carried out by \cite{helling2006dust, helling2008comparison, witte2009dust, witte2011dust, de2011influence}.
These models follow the trajectory of seed particles from the top of the atmospheres where they grow by homogeneous nucleation and descend downwards through sedimentation, acting as a cloud condensation nuclei (CCN) for a range of species.
However, the incorporation of microphysics comes at the cost of computational complexity. Comparisons with data must be conducted in a forward-modeling scheme \citep[e.g.][]{Witte2011}, as opposed to utilizing Bayesian-based inverse techniques. 

In another approach, \cite{ohno2018microphysical} utilize a microphysical model based upon a 1D Eulerian framework which calculates the vertical number distributions and mass densities of cloud particles by considering transport due to the upward motion, downward gravitational settling and eddy diffusion, and growth due to condensation and coalescence.
This approach has been used to study the effect of diffusion strength and nucleation efficiency on cloud properties \citep{ormel2019arcis} and how the porosity of cloud particle aggregates evolves in exoplanet atmospheres \citep{ohno2020clouds}.

Another microphysical model is the Community Aerosol and Radiation Model for Atmosphere (CARMA). Though it was originally intended for applications to Earth \citep{toon1979one, turco1979one}, it has subsequently been extended to a variety of planetary atmospheres \citep{toon1988multidimensional, murphy1993martian, zhao1995model, gao2014bimodal, powell2019transit}.
CARMA is a one-dimensional model that solves the discretized continuity equation for aerosol particles given vertical transport by sedimentation and diffusion as well as condensate growth and loss due to nucleation, condensation, evaporation and coagulation. Similar to the models of \cite{helling2006dust}, CARMA has also been used in a forward-modeling framework to reproduce the degree of molecular feature muting seen in exoplanet transmission spectroscopy \citep{gao2020aerosol}.

To combat the numerical expense of fully microphysical models, parametric models that prioritize a lesser number of these critical processes are proposed as a simpler yet effective method of predicting cloud profiles.
The cloud sedimentation model by \cite{ackerman2001cloud} is a widely used parametric framework that balances turbulent, upward diffusion with downward sedimentation in horizontally homogeneous clouds to obtain the one-dimensional vertical profile of condensate mass and the particle size distributions.
The vertical extent of the cloud is governed by the sedimentation efficiency $\fsed$, a constant parameter which prescribes which particles can settle out of the cloud.
Small values of $\fsed$ tend to produce thick clouds that extend vertically throughout the atmosphere whereas large values of $\fsed$ result in thin, compressed clouds.
Such parametric models have the computational advantage of producing rapid results and therefore are suitable for use in retrievals, however, with the sacrifice of microphysical detail.

This approach has been somewhat successful in interpreting the spectra of cloudy brown dwarfs \citep[e.g.][]{stephens20090} and exoplanets \citep[e.g. GJ 1214 b][]{morley2013quantitatively}.
For example, \cite{stephens20090} compared observed 5.2--14.5 $\mu$m low-resolution spectra of mid-L to mid-T dwarfs to synthetic spectra from the models of Saumon and Marley \citep{ackerman2001cloud, marley2002clouds, saumon2003non, saumon2008evolution}.
The authors found that the models agreed well with the observed spectra except for a very red L dwarf, indicating that the cloud model was inadequate for this case.
It was speculated that very red dwarfs may have particle sizes smaller than those calculated by the \citep{ackerman2001cloud} cloud model, even for small $\fsed$.
In another case, \cite{cushing2008atmospheric} performed the first fits of the 0.9--14.5 $\mu$m spectral energy distributions of nine field ultracool dwarfs with spectral types L1--T4.5 to obtain effective temperatures, gravities and cloud sedimentation efficiencies. 
The authors used the cloud modeling framework of \cite{ackerman2001cloud} to compute condensate size and vertical profiles for different values of the cloud sedimentation parameter $\fsed$.
The models fit the data reasonably well for the earliest and latest spectral types, yet for the mid to late-type L dwarfs and the early-type T dwarf the limitations of the simple cloud model resulted in prominent discrepancies.
They found that $\fsed$ varied significantly among L dwarfs with no obvious trend with spectral type.
These investigations provoke the need for further analysis into the ability of $\fsed$ to produce models that provide a good fit to the observed spectra across a wide parameter space.

The \cite{ackerman2001cloud} model has also been used to explore the impact of H$_2$O on the reflected spectra of cool giant planets \citep{macdonald2018exploring}, atmospheric retrievals \citep{molliere2019petitradtrans, mai2019exploring} and global circulation models \citep{lines2019overcast}.

Despite its physical limitations, parametric models such as \cite{ackerman2001cloud} often perform well when simulating cloud profiles, and their numerical advantage is enviable. Ideally, the physical processes that govern the magnitude of sedimentation parameter $\fsed$ could be understood within the microphysical context. For example,
\cite{gao2018sedimentation} compared the work of \cite{ackerman2001cloud} with CARMA to explore trends in $\fsed$ with eddy diffusivity ($\kzz$), gravity, material properties, and cloud formation pathways. 
\cite{gao2018sedimentation} found that $\fsed$ is sensitive to the nucleation rate of cloud particles; an attribute that is governed by material properties such as surface energy and molecular weight. 
They explain that materials with high surface energies will form fewer, larger cloud particles, corresponding to larger values of $\fsed$.
The opposite is true for materials with low surface energy. 
Furthermore, in the case of heterogeneous nucleation, the authors found that $\fsed$ depends on the CCN flux and radius. 
However, a notable feature of the comparison conducted by \cite{gao2018sedimentation} is that the cloud mass mixing ratio produced by the ``best fit'' $\fsed$ rarely agree closely with those produced by CARMA.

Therefore, in this work, we take the first step toward adding complexity to the parametric models and thus pave the way for the addition of further detail that could more closely match microphysical models.
The \cite{ackerman2001cloud} model has recently been implemented in Python package $\virga$ with several updates; the particular modifications and improvements are discussed by \cite{virga}.
We present here an extension to $\virga$ aimed at enabling greater cloud diversity and a closer comparison to the model predictions of CARMA.
Specifically, we introduce a new approach to describe sedimentation through the modification of the sedimentation efficiency $\fsed$.
In the original \cite{ackerman2001cloud} methodology, $\fsed$ is constant at every point in the atmosphere.
This is a limiting simplification; it is physically intuitive that the sedimentation would change as a function of pressure due to variable atmospheric properties.
Enforcing a constant sedimentation efficiency thereby limits the variability of cloud profiles throughout the atmosphere.

To rectify this, we have extended the definition of $\fsed$ to include altitude-dependent functions.
This development effectuates a plethora of alternative cloud profiles that are otherwise unattainable by constant values of $\fsed$.
Following the analysis conducted by \cite{gao2018sedimentation}, we compare the results of the variable $\fsed$ extension to that of CARMA to explore whether or not the greater versatility improves the agreement between microphysically and parametrically derived profiles.

We outline this work as follows:
in Section \ref{sec:models}, we describe the original \cite{ackerman2001cloud} methodology that has been updated in $\virga$.
We derive an altitude-dependent $\fsed$ function in Section \ref{sec:fsed} based upon an inferred dependence on atmospheric density. 
We discuss the free parameters of the function and describe the influence of each parameter on the behaviour of $\fsed$ throughout the atmosphere.
We then outline the impact of this new expression for $\fsed$ on the functionality of $\virga$, in particular we calculate the new solution to the diffusion-sedimentation equation that governs cloud mass and particle size distribution. 
Finally, a comparison with CARMA is conducted in Section \ref{sec:carma_comparison} to explore the influence of a variable $\fsed$ function on the model agreement. 
As this is an early stage of the variable $\fsed$ development, we compare the cloud mass mixing ratio profiles for only two of the cases from \cite{gao2018sedimentation}, namely homogeneous and heterogeneous nucleation in CARMA.
The comparison is conducted by considering the ``best fit'' $\fsed$ values for both the constant and variable options, with particular focus on whether introducing altitude-dependence has a positive advantage over results obtained using a constant $\fsed$. 
We also compare the reflected light and thermal spectra consequential of the constant and variable $\fsed$ formulations.
We conclude our findings and outline future work in Section \ref{sec:conclusions}.

\section{Ackerman \& Marley and VIRGA model}
\label{sec:models}

The methodology of \cite{ackerman2001cloud} has been implemented in new Python package $\virga$ by \cite{virga}. 
This model solves for horizontally homogeneous clouds, where the vertical extent of the cloud is governed by a balance between upward turbulent diffusion and downwards sedimentation.
The diffusion-sedimentation equation is given by
\begin{equation}
    -\kzz\frac{\partial q_t}{\partial z} - \fsed \w q_c = 0,
    \label{eq:diff-sed-eqn}
\end{equation}
where $\kzz$ is the eddy diffusion coefficient, $\fsed$ is the sedimentation efficiency, $q_t$ is the total (vapour + condensate) mass mixing ratio, $q_c$ is the condensate mass mixing ratio, $z$ is the altitude and $\w$ is the mean upward velocity\footnote{\cite{ackerman2001cloud} refer to $\w$ as the convective velocity, though it is also being applied in the radiative layer. Therefore we introduce this change in terminology in this work.}.
Equation \eqref{eq:diff-sed-eqn} is solved for every condensible species independently, therefore we neglect any microphysical interactions between clouds.

Following the methodology of \cite{ackerman2001cloud}, the parameter $\fsed$ is a user-defined constant, primarily to ensure an analytical solution to \eqref{eq:diff-sed-eqn}. 
To solve \eqref{eq:diff-sed-eqn}, the atmosphere is divided into vertically homogeneous layers and solved from the bottom-up (high to low pressure). 
We assume a constant mixing ratio of saturated vapour $\qvs$ within each layer and require all excess vapour to condense.
The boundary condition for the bottom-most layer of the cloud deck is a user-defined mixing ratio, usually derived from equilibrium chemistry calculations \citep[e.g.][]{visscher2006atmospheric}. 
For each successive layer after the bottom-most layer, the lower boundary condition $q_\text{below}$ is equal to the mixing ratio of the layer below.
Rescaling the altitude within each layer to vary between $\hat{z}=0$ and $\hat{z}=\mathrm{d}z$, where $\mathrm{d}z$ is the width of the layer, the solution to \eqref{eq:diff-sed-eqn} at the top of a layer for constant $\fsed$ is given by
\begin{equation}
    q_t(\mathrm{d}z) = \qvs + (q_\text{below} - \qvs)\exp\left(-\frac{\fsed\mathrm{d}z}{L}\right).
\end{equation}
The parameters necessary to prescribe a lognormal distribution of condensate particle sizes are then deduced within each layer.

The analytical capability of $\virga$ is vital in retaining its superior numerical efficiency.
However, a constant $\fsed$ doesn't exclusively enable this; we simply require an explicit anti-derivative for $\fsed$.
Therefore we can introduce altitude-variation into $\fsed$ without sacrificing the analytical capability of the model.

\section{Derivation of variable $\fsed$}
\label{sec:fsed}
The challenge is to define a sedimentation efficiency that is reflective of the physical phenomena we expect within a cloud.
From \eqref{eq:diff-sed-eqn}, small values of $\fsed$ represent the domination of upward diffusion over downward sedimentation, resulting in vertically thick clouds that extend throughout the atmosphere.
On the other hand for large values of $\fsed$, sedimentation is the dominant process, thus clouds are thinner and deplete quickly.
Defining a variable function for $\fsed$ requires careful consideration into the microphysical processes governing the particle size distribution and particle density at each point in the atmosphere.
As a first attempt, we choose to consider the atmospheric density dependence of $\fsed$ in order to derive an altitude-dependent function.
We emphasize that the $\fsed$ expression studied in this work is only one such possible parameterization with the purpose of illustrating the extended abilities of $\virga$.
We do not argue that this is the optimal or most physically meaningful expression for $\fsed$.

Recall that $\fsed$ is defined as the ratio of sedimentation velocity $\vf$ to mean upward velocity $\w$, namely
\begin{equation}
	\fsed = \frac{\vf}{\w} = \frac{\vf L}{\kzz},
	\label{eq:fsed_ratio}
\end{equation}
where $L$ is the mixing length and
\begin{equation}
    \vf = \sqrt{\frac{2mg}{\rho_a A C_d}}, \qquad \kzz = \frac{H}{3}\left(\frac{L}{H}\right)^{\frac{4}{3}}\left(\frac{RF}{\mu\rho_a c_p}\right)^{\frac{1}{3}},
    \label{eq:vf_kzz}
\end{equation}
where $m$ is the mass of the falling particle, $g$ is its acceleration due to gravity, $\rho_a$ is the atmospheric density, $A$ is the particle's projected area, $C_d$ is the drag coefficient, H is the scale-height, $R$ is the universal gas constant, $F$ is the heat flux, $\mu$ is the atmospheric molecular weight and $c_p$ is the specific heat of the atmosphere at constant pressure \citep{ackerman2001cloud}.
Assuming $H=L$ and considering only the dependence of $\vf$ and $\kzz$ on atmospheric density, we can combine equations \eqref{eq:fsed_ratio} and \eqref{eq:vf_kzz} in order to deduce a parameterization of $\fsed$ in atmospheric density:
\begin{equation}
	\fsed = C\rho_a^{-\frac{1}{6}},
	\label{eq:fs_rho}
\end{equation}
where $C$ is a constant of proportionality. 
For an isothermal atmosphere in hydrostatic equilibrium, the atmospheric density is related to the altitude $z$ as 
\begin{equation}
	\rho_a = \rho_0\exp\left(-\frac{z}{H}\right),
	\label{eq:rho}
\end{equation}
where $\rho_0$ is the atmospheric density at $z=0$.  
Combining equations \eqref{eq:fs_rho} and \eqref{eq:rho}, we obtain the altitude-dependent expression for $\fsed$
\begin{equation}
	\fsed(z) = \alpha\exp\left(\frac{z}{6H_0}\right),
	\label{eq:fsed1}
\end{equation}
where $\alpha$ is a constant of proportionality. 
We note that because we will use $\fsed$ in a non-isothermal atmosphere and scale height varies with temperature, for the purpose of this analysis we fix the scale-height in our expression for $\fsed$ to be the scale-height $H_0$ at a pressure of 1 bar. 

It is clear that our expression for $\fsed$ is unbounded as altitude increases, in particular, 
\begin{equation}
	\lim_{z\rightarrow\infty}\fsed(z)=\infty.
\end{equation}
We wish to rescale our expression to ensure finite values throughout the atmosphere and increase the lower bound.
Previous work has shown that $\fsed$ can be as small as 0.01 for super Earths \citep[e.g.][]{morley2015thermal} or 0.1 for hot Jupiters \citep[e.g.][]{webber2015effect}. 
Models with $\fsed \ll 0.01$ result in particle sizes that have effective radius that approach the Van der Waals derived radii of the gas particles. Therefore, particles of that size are typically assumed to be evaporated back into the gas. 
Our model is robust against this as it disallows computations if the mean particle size minus 0.75$\sigma$, where $\sigma$ is the standard deviation of the lognormal distribution, is smaller than the pre-computed grid of Mie parameters.
We constrain \eqref{eq:fsed1} with input parameter $\epsilon$ to ensure that $\fsed\geq\epsilon$ (where $\epsilon\geq 0.01$) to mitigate calculation of particle sizes that are too small:
\begin{align}
	\fsed(z) &= \alpha\exp\left(\frac{z}{6 H_0}\right) + \epsilon.
	\label{eq:fsed2}
\end{align}
We also normalise $\fsed$ such that it takes on a value $\alpha+\epsilon$ at the user-defined value of $\zT$ (where $P(z=\zT)\le10^{-4}$bar):
\begin{equation}
	\fsed(z) = \alpha\exp\left(\frac{z-\zT}{6H_0}\right) + \epsilon.
	\label{eq:fsed3}
\end{equation}
Our expression for $\fsed$ now satisfies the following constraints:
\begin{equation}
	\fsed(\zT) = \alpha+\epsilon,\qquad\fsed(z) \geq \epsilon,\,\,\forall z\geq 0.
	\label{eq:og_constraints}
\end{equation}
Finally, we introduce scaling parameter $\beta$ in the denominator of the exponential. 
The purpose of this parameter is to control the rate of change of $\fsed$ with respect to altitude, which is sensitive to the assumptions we made when writing $\fsed$ as a function of $\rho_a$ in \eqref{eq:fs_rho} and our limitation of scale-height to the constant value $H_0$: 
\begin{equation}
	\fsed(z) = \alpha\exp\left(\frac{z-\zT}{6\beta H_0}\right) + \epsilon.
	\label{eq:fsed}
\end{equation}
Given equation \eqref{eq:fsed}, large values of $\beta$ will result in near constant $\fsed$ profiles of approximate magnitude $\alpha+\epsilon$, while small values of $\beta$ produces an $\fsed$ profile that is smaller at depth and higher towards lower pressures.

This formalism neglects the vertical dependence of the drag coefficient and the mass-to-area ratio for the cloud particles that play a central role in controlling cloud properties. 
However, in this proof-of-concept work, we aim to take the first step in expanding the \cite{ackerman2001cloud} methodology. 
We reiterate that this is only one example of a variable $\fsed$ function and we are not restricted to this choice;
\texttt{virga} could support any variable $\fsed$ function that has an explicit anti-derivative.
From the fundamental theorem of calculus, anti-derivatives exist for every continuous, real-valued function. 
Straightforward examples of such functional forms include polynomials, logarithms and logistic curves. 
By considering the atmospheric density as the defining parameter for the behaviour of $\fsed$ throughout the atmosphere we arrived at an expression that increases from the bottom to the top of the atmosphere. 
As a check, we considered an expression that exhibited the opposite behaviour, that is, one that decreases from the bottom to the top of the atmosphere.
By using this decreasing expression in the CARMA comparison outlined in Section \ref{sec:carma_comparison}, we found that the $\fsed$ function that allowed the optimal fit between $\virga$ and CARMA is essentially constant, therefore, the capacity for such a variation throughout the atmosphere was irrelevant.
In contrast, the increasing $\fsed$ expression discussed throughout the paper enables significantly better agreement between models.

Alternative choices for $\fsed$ may be more appropriate for other condensates, atmospheric properties and celestial bodies, therefore, adequate investigation into different functional forms of $\fsed$ is an important next step.
Equation \eqref{eq:fsed} is the $\fsed$ function that we will use throughout this analysis.

\subsection{Effect of free parameters on $\fsed$}
\begin{figure*}[b!]`
\centering
	\gridline{\fig{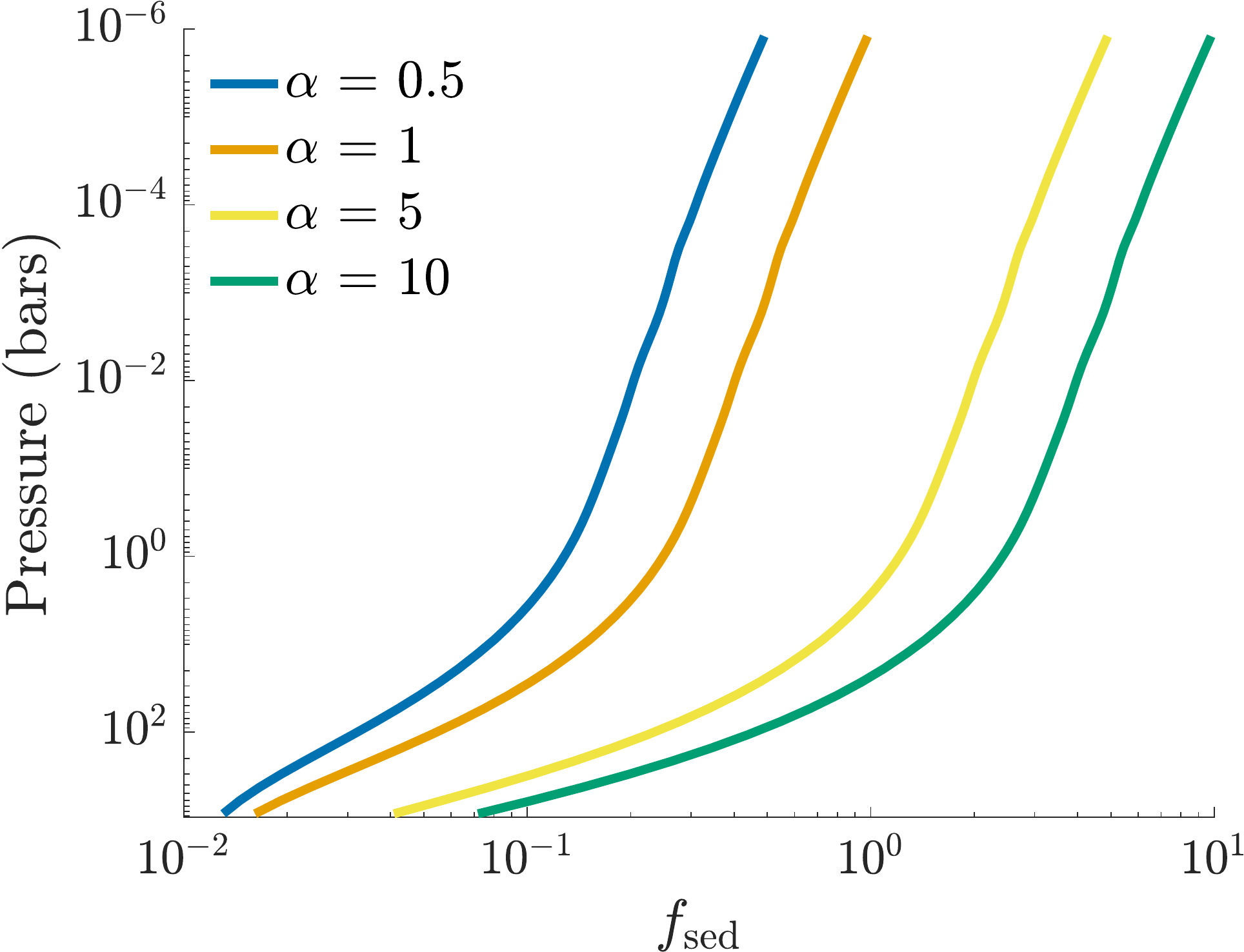}{0.4\textwidth}{(a) $\beta=1$.}
          			\fig{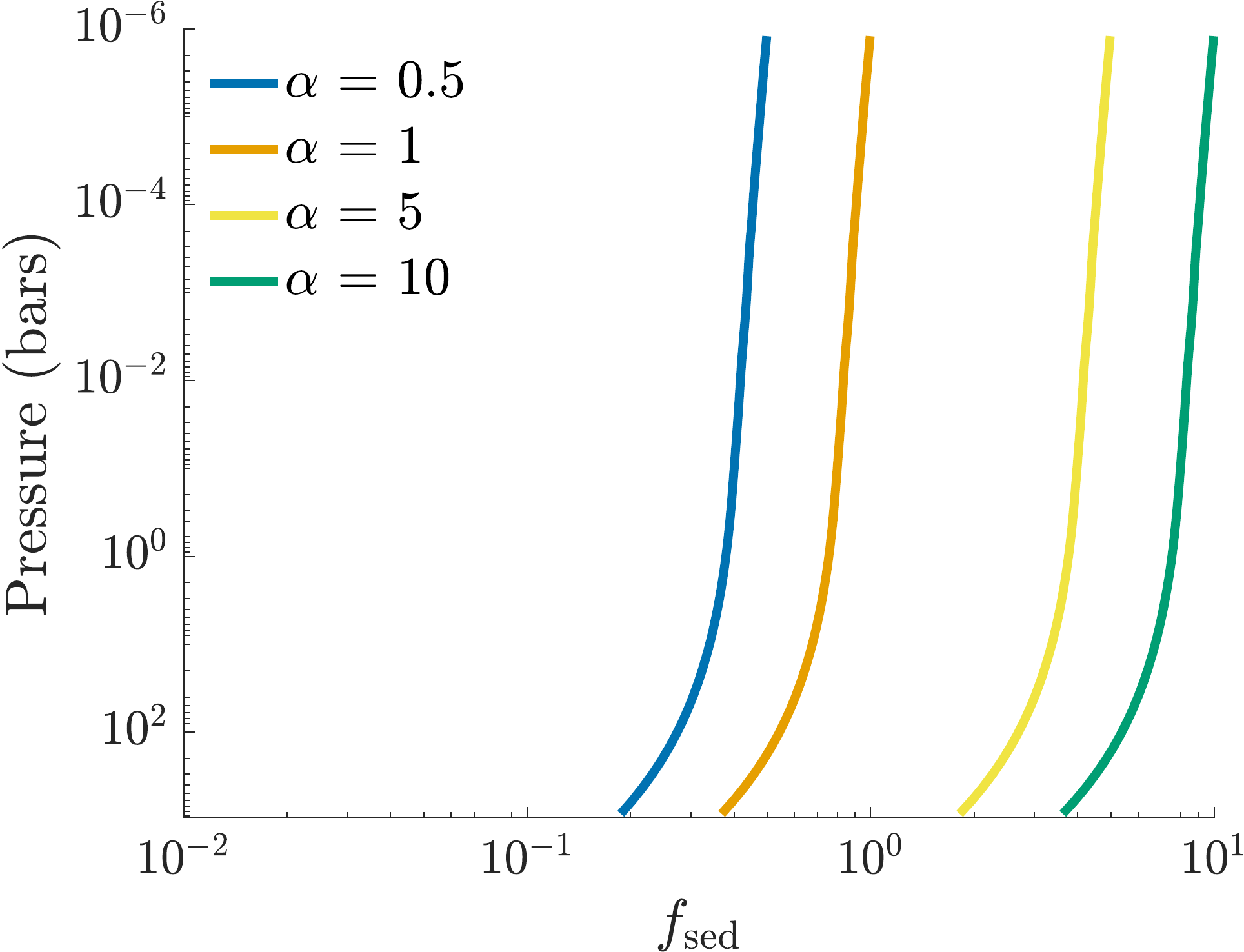}{0.4\textwidth}{(b) $\beta=5$.}
          }
	\gridline{\fig{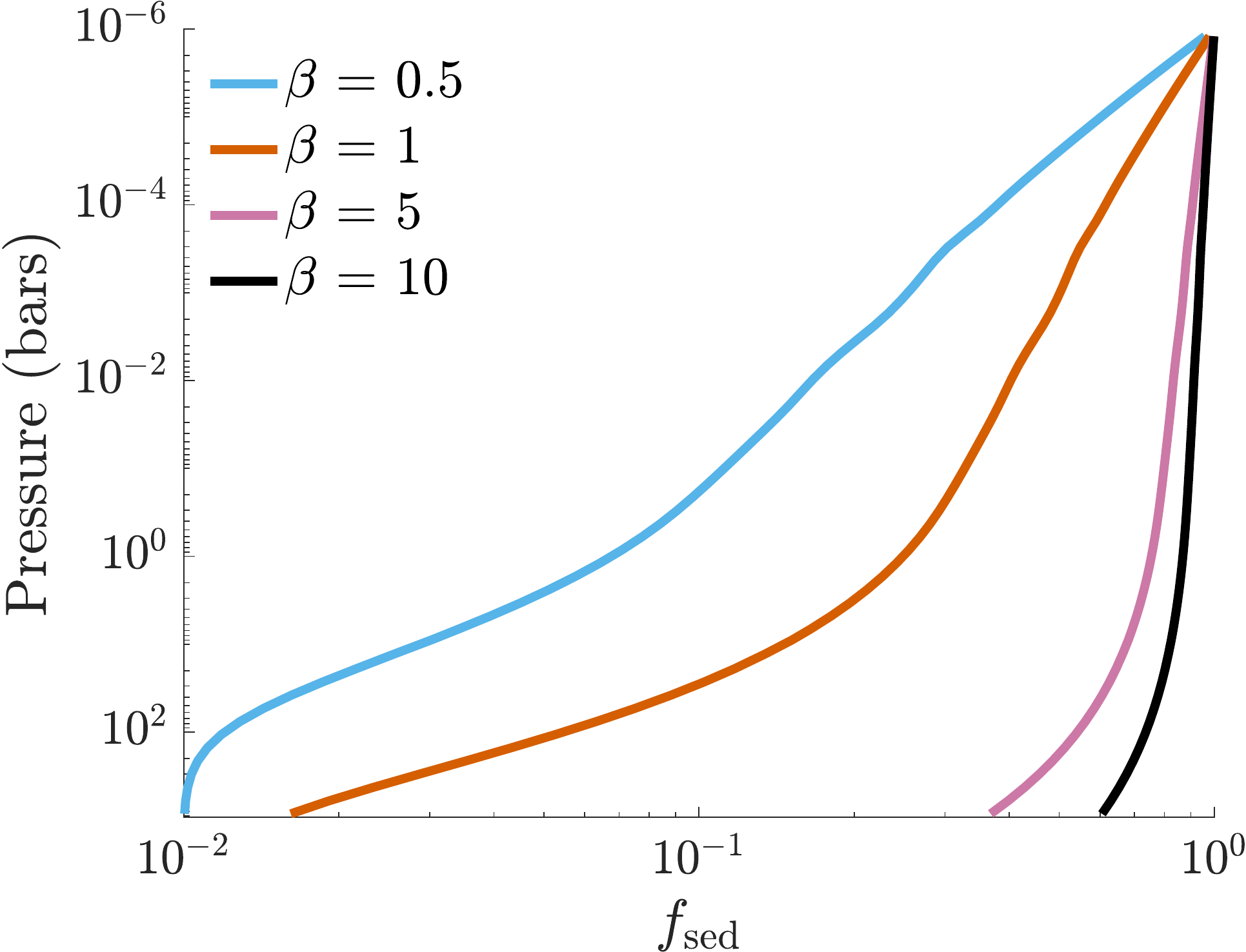}{0.4\textwidth}{(c) $\alpha=1$.}
         			 \fig{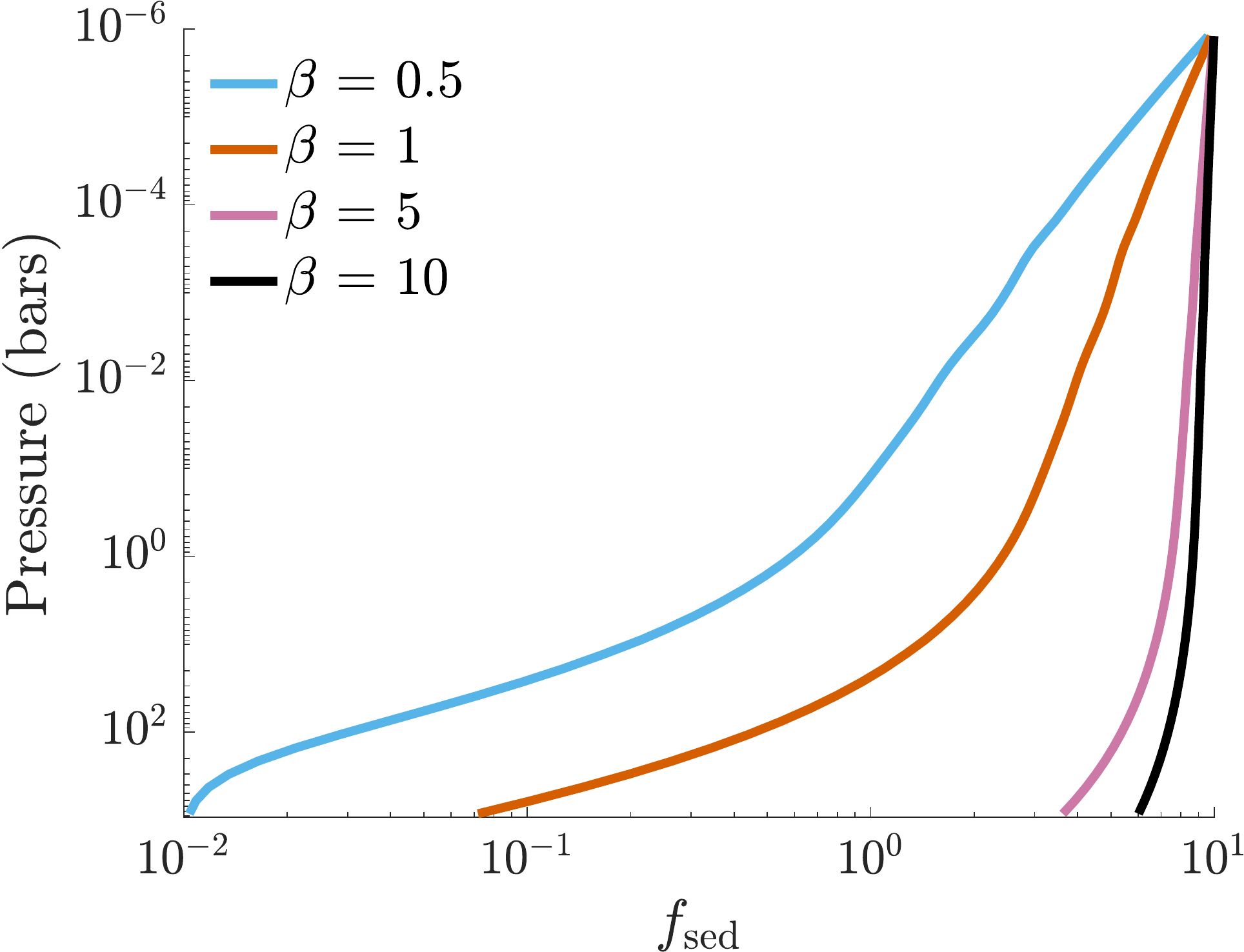}{0.4\textwidth}{(d) $\alpha=10$.}
          }
	\caption{We consider different values of $\alpha$ and $\beta$ in our expression for $\fsed$ \eqref{eq:fsed} to study the variation of $\fsed$ with pressure. 	We fix $\epsilon=10^{-2}$ and $P(z=z_T)=10^{-6}$ bar. Increasing $\alpha$ increases the maximum value of $\fsed$ and also results in a nearly uniform shift increase at every point in the atmosphere. Small values of $\beta$ force $\fsed$ to vary gradually over the extent of the atmosphere whereas larger values result in a steep slope with little variation.}
	\label{fig:fsed}
\end{figure*}

To analyse the effect of the choice of $\alpha$ and $\beta$ on the behaviour of $\fsed$ throughout the atmosphere we plot pressure against $\fsed$ in Figure \ref{fig:fsed} for a variety of parameter values.
We see from Figures \ref{fig:fsed}(a) and \ref{fig:fsed}(b) that the shape of the $\fsed$ curve is almost unchanged as we vary $\alpha$ but the entire curve undergoes a nearly uniform shift increase as $\alpha$ is increased. 
The value of $\fsed=\alpha+\epsilon$ is obtained at $z_T$, close to the top of the atmosphere.
Exponential parameter $\beta$ influences the rate of change of $\fsed$ with pressure, as evidenced by Figures \ref{fig:fsed}(c) and \ref{fig:fsed}(d) .
Smaller values of $\beta$ result in a gradual increase from the minimum to maximum values of $\fsed$ over the extent of the atmosphere, whereas larger values of $\beta$ produce a sharp increase leaving little room for $\fsed$ variation within the atmosphere.
Unless otherwise stated, we take $\epsilon=10^{-2}$ and $z_T$ to be the top of the atmosphere in this analysis.
 
\subsection{New solution to the diffusion-sedimentation equation}
\label{sec:new_soln}
As with the original methodology outlined in Section \ref{sec:models}, we compute the vertical distributions of condensate and vapour by proceeding upwards from the subcloud conditions, using the subcloud mixing ratio $q_\text{below}$ as the lower boundary condition. 
We solve the diffusion-sedimentation equation \eqref{eq:diff-sed-eqn} for variable $\fsed$ of the form \eqref{eq:fsed3} at the boundaries between each layer. 
By rescaling the problem such that the altitude within each layer varies between $\hat{z}=0$ and $\hat{z}=\mathrm{d}z$, we obtain the new solution for total mixing ratio to be
\begin{align}
    q_t(\mathrm{d}z) &= \qvs + (q_\text{below} - \qvs)\exp\left(\hat\alpha \left[\exp\left(\frac{\mathrm{d}z}{6\beta H_0}\right)-1\right] - \frac{\epsilon \mathrm{d}z}{L}\right)
\end{align}
where $\hat{\alpha}=-\frac{6\alpha\beta H_0}{L}\exp\left(\frac{\zB-\zT}{6\beta H_0}\right)$ and $\zB$ is the altitude at the bottom of the layer.

\section{Comparison with CARMA model}
Following the goals of \cite{gao2018sedimentation}, we compare the cloud mass mixing ratios for KCl clouds computed by CARMA and $\virga$ as a function of pressure. Specifically, we focus on two CARMA cases: 1) homogeneous nucleation with varying surface energy of KCl and 2) heterogeneous nucleation with varying downward flux of CCN.
We note that, unlike CARMA,  $\virga$ does not explicitly incorporate any nucleation calculations in its methodology.
By conducting the comparison for different nucleation mechanisms in CARMA, we are exploring whether $\virga$ can produce a cloud mass mixing ratio profile that resembles that of CARMA by varying $\fsed$ with altitude instead of directly including the microphysical intricacies of nucleation processes.

In order to have an effective comparison, it is critical to have identical inputs: pressure-temperature profiles, lower-boundary condensate mixing ratios, saturation vapor pressure, and $\kzz$ profiles. Therefore, we adopt the constant $\kzz$ profiles of $10^7$ and $10^8$~cm$^{2}$s$^{-1}$ as in \citet{gao2018sedimentation}. 
For the pressure-temperature models, those of \cite{gao2018sedimentation} were extracted from the Sonora model grid \citep{marley2018sonora}. Briefly, these temperature-pressure profiles are computed from a radiative-convective-thermochemical equilibrium model \citep{marley1999thermal} for a range of effective temperatures, and gravities at a fixed solar metallicity. The Sonora grid also assumes no external insolation, and thus are only suitable for brown dwarfs or cool giant planets.
The profile explored in \citet{gao2018sedimentation} has an effective temperature of 400~K and $\log g=5.25$ ($g$ in cgs units). 
We reiterate the note of \cite{gao2018sedimentation} that objects with an effective temperature of 400~K and $\log g=5.25$ do not yet exist because the large mass associated with such objects requires a timescale longer than the age of the universe to cool to such low temperatures.
As the present paper is intended as a proof-of-concept of the computational versatility permitted by variable $\fsed$ profiles rather than trying to predict actual clouds on real objects, such a test case is acceptable.

We use a fixed lower boundary mixing ratio of 0.22~ppmv for CARMA for KCl vapor. 
The saturation vapor pressure estimated by \cite{morley2012neglected} is given as
\begin{equation}
    \log p_s^\text{KCl} = 7.611 - \frac{11382}{T},    
\end{equation}
where $p_s^\text{KCl}$ is measured in bars and $T$ in K.
In CARMA, upon reaching saturation, KCl nucleates homogeneously to form a cloud deck and proceeds to evolve by condensation and evaporation while being transported by sedimentation and diffusion.
The homogeneous nucleation cases are subject to a zero-flux upper boundary condition whereas a finite flux of condensation nuclei is enforced for the heterogeneous nucleation cases.

We conduct the comparisons between CARMA and $\virga$ by finding the optimal constant $\fsed$ value and variable $\fsed$ profile that ``best fit'' the cloud mass mixing ratio distributions computed by CARMA.
\cite{gao2018sedimentation} defined the best-fit $\fsed$ to be that which minimizes the difference in the pressure level where each model reaches a cumulative optical depth of 0.1, denoted $P_{0.1}$.
We opt for a different definition that better incorporates the vertically-dependent profile. We instead choose the optimal $\fsed$ parameters to be those that minimize the distance (i.e. chi-square value) between the mass mixing ratio profiles of the two models in the region of  $\log P_{0.1}\pm1$, where $P_{0.1}$ is that of CARMA. 
Constraining the pressure region over which we compare the profiles is preferable to fitting throughout the entire vertical extent because the upper regions of the cloud, which are more diffuse, and lower regions which are very optically thick, do not impact transmission, reflection and emission observations. Our ultimate goal is to better reproduce spectral observations of exoplanets and brown dwarfs, thus, it is less critical that the two models agree here.
Note that to evaluate the cumulative optical depth, \cite{gao2018sedimentation} ignored wavelength dependence and instead used the conservative geometric scattering approximation with an extinction coefficient of 2.

A key difference in the modeling of CARMA and $\virga$ is the treatment of condensation; $\virga$ assumes that all vapour in excess of saturation condenses, whereas condensation in CARMA is limited by the nucleation energy barrier.
The nucleation energy barrier is the energy needed for a species to undergo nucleation.
As a consequence, $\virga$ will generally predict a greater cloud condensate mass mixing ratio than CARMA.
In order to compare the two models under near-identical conditions, we limit the mass mixing ratio of KCl in $\virga$ to simulate the effect of a nucleation energy barrier similar to that of CARMA.
In particular, we force the lower boundary mixing ratio of KCl $q_v^\mathrm{KCl}$ in $\virga$ to be the maximum value of the condensate mass mixing ratio calculated by CARMA, $q_{c,\text{CAR}}^\mathrm{KCl}(z)$, throughout the atmosphere, namely
\begin{equation}
    q_v^\mathrm{KCl} = \max_{\forall z}(q_{c,\text{CAR}}^\mathrm{KCl}(z)).
\end{equation}

The lower boundary mass mixing ratio is informed in both CARMA and \cite{ackerman2001cloud} (and hence $\virga$) from chemical equilibrium models \citep[e.g.][]{visscher2010atmospheric}. The lower boundary mass mixing ratio in CARMA can be further reduced by the surface energy of the condensate, which is a well-documented and predictable effect. Though $\virga$ does not include this, the noted benefit of it is not it's self-consistency, but it's computational efficiency that allows it to be run within a retrieval framework, grid framework, or iterative climate code.
In these cases where $q_c$ is unknown, $q_c$ could be treated as a free-parameter if the intent was to study deviations from chemical equilibrium.
Therefore, our variable $\fsed$ framework is still useful in the case of unknown $q_c$.

\subsection{Heterogeneous nucleation}
\label{sec:carma_comparison}
 \begin{figure*}[t!]
	\centering
	\gridline{\fig{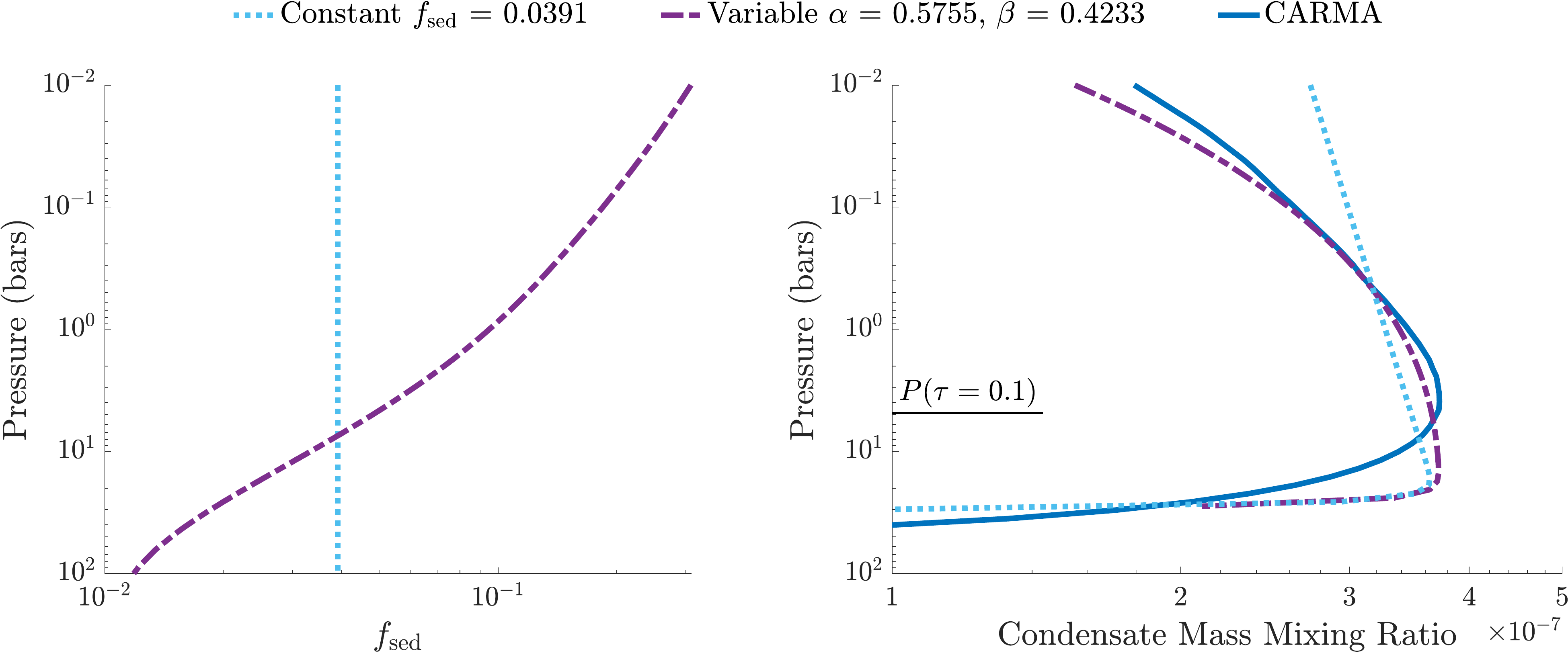}{0.9\textwidth}
					{(a) CCN flux = 100 cm$^{-2}$s$^{-1}$.}
          }\vspace{0.5cm}
	\gridline{\fig{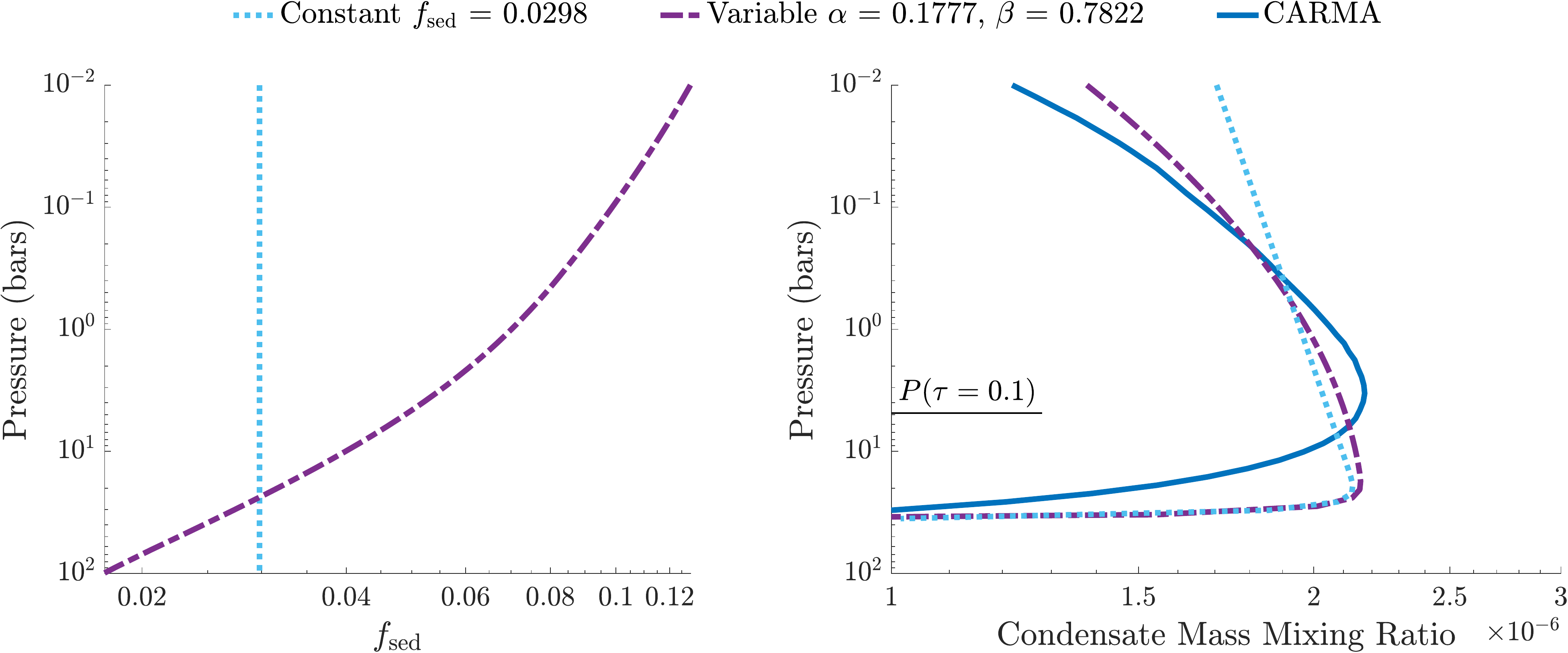}{0.9\textwidth}
				{(b) CCN flux = 1000 cm$^{-2}$s$^{-1}$.}
          }
	\caption{Cloud mass mixing ratio $q_c$ with pressure computed by CARMA with heterogeneous nucleation (solid) and \texttt{virga} using the best-fit constant $\fsed$ (dotted) and variable $\fsed$ (dash-dot). For CARMA, the downwards flux of cloud condensation nuclei (CCN) is (a) 100~cm$^{-2}$s$^{-1}$ and (b) 1000~cm$^{-2}$s$^{-1}$, while the CCN radius is fixed to 1~nm. For both models, $\kzz=10^8$~cm$^2$~s$^{-1}$ and $\log g=5.25$. The pressure levels where the cumulative optical depth in CARMA reaches 0.1 from above are shown as horizontal solid lines, and the best-fit $\fsed$ is chosen to minimize the difference between the cloud mass mixing ratio profiles of the two models within one order of magnitude either side of this pressure level. }
	\label{fig:fig7_qc_comparison}
\end{figure*}
\cite{gao2018sedimentation} recognized that the cloud distribution produced by heterogeneous nucleation is dependent on the downward flux of cloud condensation nuclei (CCN), noting that the cloud mass density scales approximately linearly in log-space with nuclei flux.
The ``best-fit'' A\&M profile failed to capture the behaviour of the CARMA profile.
The authors consider three flux values, namely 10, 100 and 1000~cm$^{-2}$s$^{-1}$ with condensation nuclei radii 0.1~nm.
CARMA assumes that the condensation nuclei are composed of meteoritic  dust with small contact angle 0.1$^\circ$.
In this work, we use the CARMA profile with 100 and 1000~cm$^{-2}$s$^{-1}$ downward flux as the representative cases for heterogeneous nucleation.
The comparison between CARMA and \texttt{virga} with both best-fit constant and variable $\fsed$ values is depicted in Figure \ref{fig:fig7_qc_comparison}. 
The best-fit constant value for 100~cm$^{-2}$s$^{-1}$ downward flux is $\fsed=0.0391$ and the best-fit variable parameters are $\alpha=0.5755$, $\beta=0.4233$ (Figure \ref{fig:fig7_qc_comparison}(a)) and the corresponding values for 1000~cm$^{-2}$s$^{-1}$ are $\fsed=0.0298$, $\alpha=0.1777$ and $\beta=0.7822$ (Figure \ref{fig:fig7_qc_comparison}(b)).

In both cases, the cloud profile produced by \texttt{virga} with the variable $\fsed$ function agrees significantly better with the CARMA profile than the constant $\fsed$ result.
In particular, the variable $\fsed$ profile better captures the curvature of the CARMA profile. 
We conjecture that the reason for this improvement is that this particular heterogeneous nucleation case is limited by the number of nucleation centers imposed by the relatively low CCN flux. Figure 7 of \cite{gao2018sedimentation} shows that none of the heterogeneously nucleated cloud mass mixing ratio profiles approach the maximum mass mixing ratio possible, a fully mixed cloud profile. This suggests that only a small fraction of the available condensate vapor were able to nucleate into cloud particles. Under such conditions, the CCN number density profile is critical for determining the cloud mass mixing ratio profile. As such, a lower CCN number density at the top of the cloud due to higher CCN sedimentation velocities coupled with lower nucleation rates from typically lower temperatures  (compared to near the cloud base) could result in low cloud particle number densities. These particles can then grow to relatively large sizes via condensation, since the available vapor is shared among only a few particles. 
Ultimately, this results in an $\fsed$ that increases with altitude. 
Based upon this observation, we speculate that a cloud with a mass mixing ratio profile limited by the abundance of CCN may be better described with a variable $\fsed$ instead of a constant one, however, further analysis is needed to draw a definitive conclusion.
We also note that an $\fsed$ increasing with altitude is not a unique characteristic of clouds formed via heterogeneous nucleation.
More work is needed to rigorously explore which cloud formation mechanisms are best represented by variable $\fsed$, along with the optimal functional form of such expressions.

\subsection{Homogeneous nucleation}
 \begin{figure*}[t!]
	\centering
	\gridline{\fig{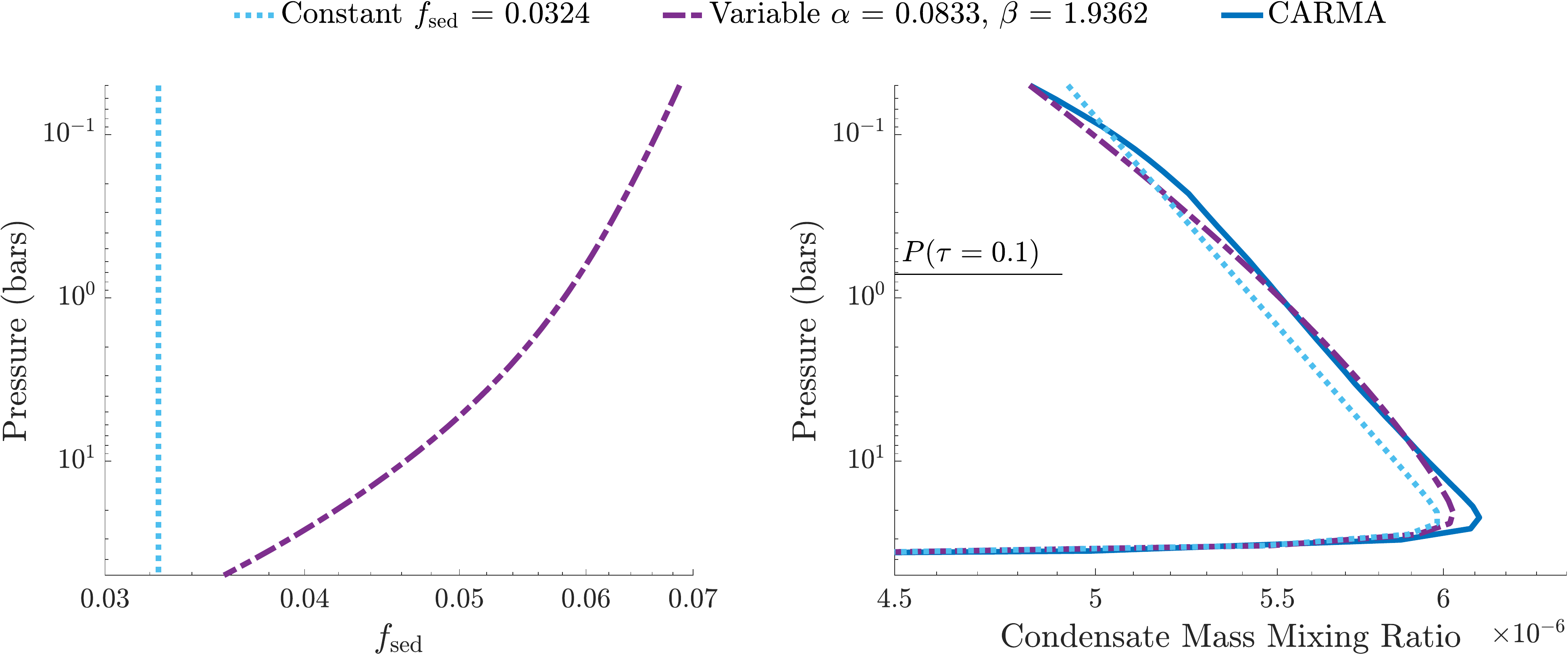}{0.9\textwidth}
					{(a) Surface energy = $\sigma_0^\text{KCl}/4.$}
          }\vspace{0.5cm}
	\gridline{\fig{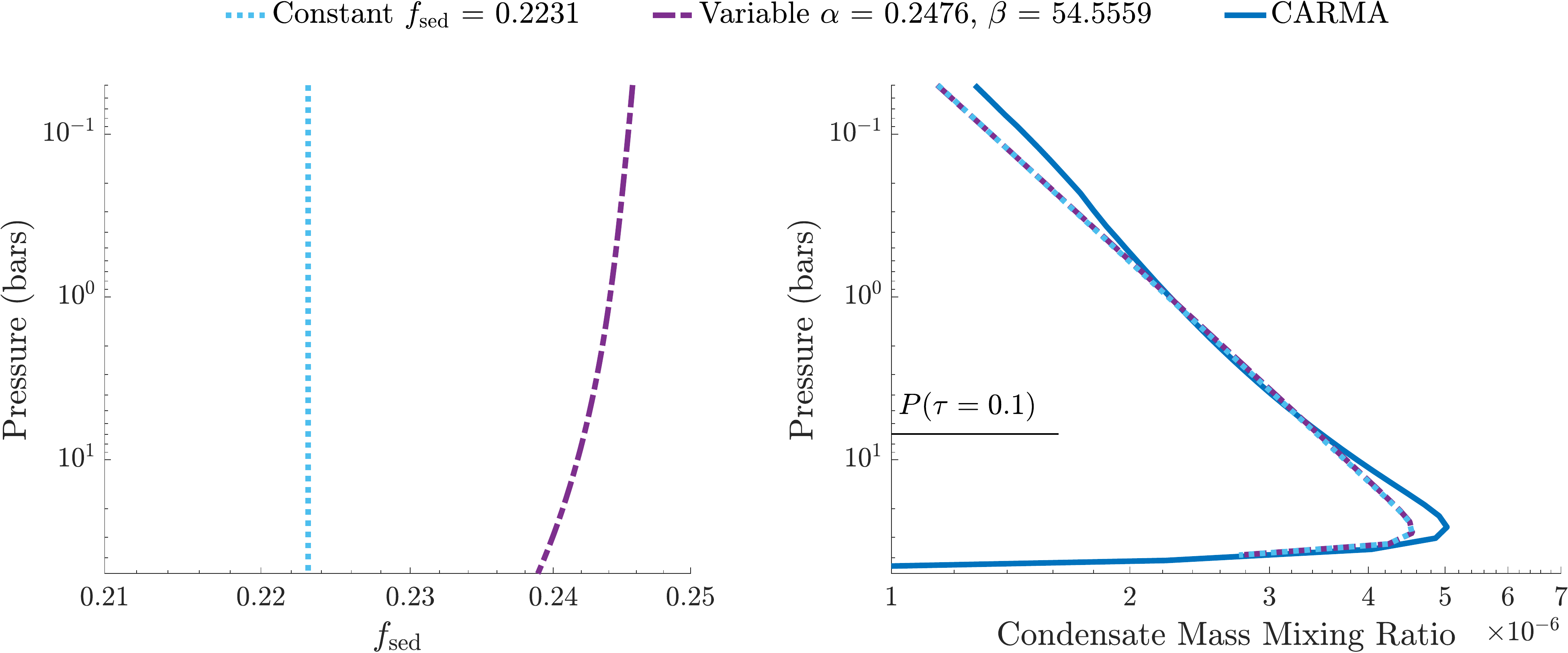}{0.9\textwidth}
				{(b) Surface energy = $2\sigma_0^\text{KCl}.$}
          }
	\caption{Cloud mass mixing ratio computed by CARMA with homogeneous nucleation (solid) compared to that of \texttt{virga} with best-fit constant $\fsed$ (dotted) and variable $\fsed$ (dash-dot) where KCl's surface energy $\sigma_0$ is altered. Diffusion coefficient $\kzz$ is $10^7$~cm$^2$s$^{-1}$ and $\log g$ is fixed at 5.25. The pressure levels where the cumulative optical depth in CARMA reaches 0.1 from above are shown as horizontal solid lines, and the best-fit $\fsed$ is chosen to minimize the difference between the cloud mass mixing ratio profiles within one order of magnitude either side of this pressure level. }
	\label{fig:fig4_qc_comparison}
\end{figure*}
In another case study, \cite{gao2018sedimentation} considered homogeneous nucleation and for different surface energies of KCl. 
The authors reported that the ``best-fit'' A\&M cloud mass mixing ratio still differed significantly from the CARMA profile.
They attributed this to a strong dependence of the cloud distribution on condensate material properties, captured primarily through the homogeneous nucleation rate, which is not explicitly included in A\&M or \texttt{virga}.

To analyze the impact of surface energy on the cloud mass mixing ratio, \cite{gao2018sedimentation} altered the surface energy of KCl (given as 160.4 - 0.07$T$($^\circ$C)) by decreasing it by factors of 2 and 4, and increasing it by factors of 2, 3 and 4.
We present the cloud mass mixing ratios produced by CARMA for $0.25\times\sigma_0^\text{KCl}$ in Figure \ref{fig:fig4_qc_comparison}(a) and that for $2\times\sigma_0^\text{KCl}$ in Figure \ref{fig:fig4_qc_comparison}(b), compared with \texttt{virga} subject to both the best-fit constant and variable $\fsed$ profiles for each case.
The best-fit constant value for the former case is $\fsed=0.0324$, and the best-fit variable parameters are $\alpha=0.0833$, $\beta=1.9362$. The equivalent parameters for the latter case are constant $\fsed=0.2231$, $\alpha=0.2476$ and $\beta=54.556$,
The high $\beta$ value in the second case indicates a nearly-constant profile at $\alpha=0.2476$, which amounts to a significantly similar profile to the constant $\fsed=0.2231$.

In contrast to the heterogeneous nucleation comparison in Figure \ref{fig:fig7_qc_comparison}, the difference between the best-fit constant and variable $\fsed$ profiles is negligible, in particular for the $2\times\sigma_0^\text{KCl}$ case.
This implies that there is little benefit to the variable $\fsed$ expression given by \eqref{eq:fsed} for homogeneous nucleation. 
Unlike heterogeneous nucleation, homogeneous nucleation is not dependent on CCNs, and is limited purely by the nucleation energy barrier, which is a function of material properties and supersaturation. KCl has a low nucleation energy barrier thanks to its low surface energy \citep{lee2018nucleation}, and thus nucleation proceeds quickly. As a result, the cloud distribution is not nucleation limited, and is instead primarily shaped by growth by condensation and transport via sedimentation and turbulent diffusion \citep{gao2018gj1214b}, the latter processes being well captured by \cite{ackerman2001cloud} with constant $\fsed$. Though it is possible that variable $\fsed$ may be preferred for homogeneously nucleated clouds made up of materials with higher nucleation energy barriers, such materials are unlikely to form optically thick clouds via homogeneous nucleation. 

\subsection{Particle size distribution}
\begin{figure*}[b!]
    \centering
    \includegraphics[width=\textwidth]{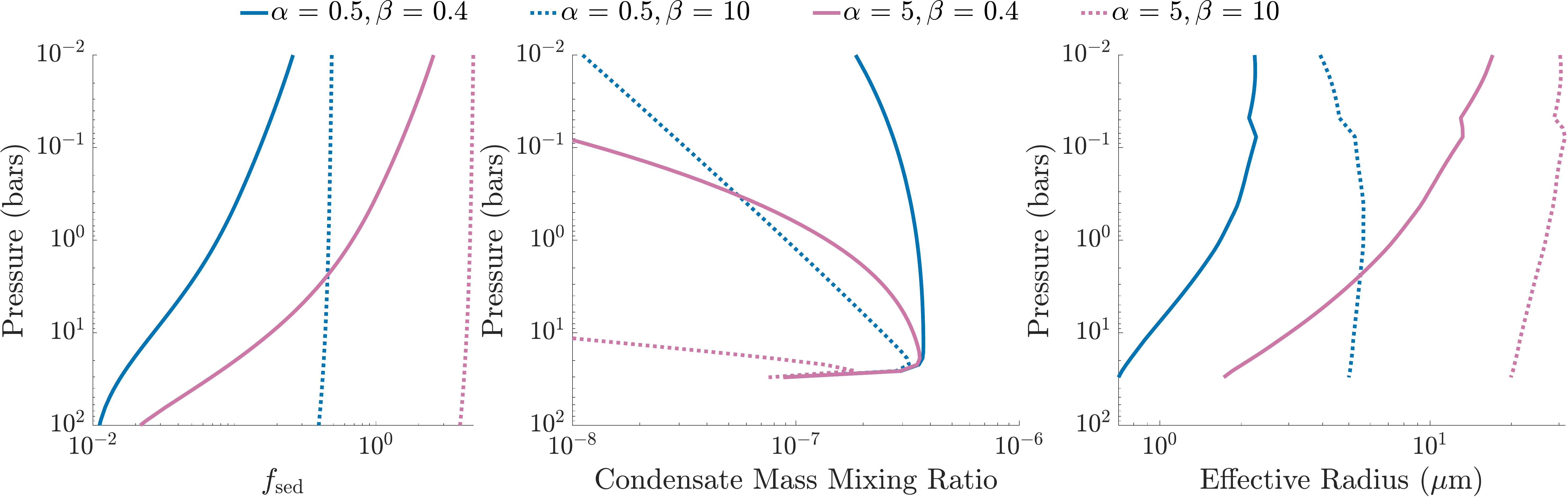}
    \caption{Influence of variable $\fsed$ \eqref{eq:fsed} parameters $\alpha$ and $\beta$ on the condensate mass mixing ratio (middle pane) and effective particle radius (rightmost pane). The blue lines represent $\alpha=0.5$ whereas the pink represent $\alpha=5$. The solid and dotted lines depict $\beta=0.4$ and $\beta=10$ respectively. We note that for large $\beta=10$ the $\fsed$ profile is approximately constant at $\fsed=\alpha$.}
    \label{fig:effect_of_params}
\end{figure*}

The size distribution of condensate particles is a critical component for determining the scattering properties of clouds in atmospheres.
Condensate particle radii commonly follow a bimodal number distributions, where a mode due to condensational growth at modest supersaturations arises $\sim 10\mu$m along with a precipitation mode at larger radii \citep{ackerman2001cloud}. 
However, for analytical feasibility, $\virga$, following the \citep{ackerman2001cloud} methodology,  makes no attempt to model the intricacies of cloud processes necessary to yield such a bimodal distribution.
Instead, $\virga$ prescribes a lognormal distribution of condensate particles within each layer.
Conversely, CARMA resolves the particle size distribution using mass bins rather than assuming any size distribution shape \citep{gao2018sedimentation}.
The disparity between how the size distributions are calculated naturally leads to discrepancies between the effective radii of particles calculated by the two models.
\cite{gao2018sedimentation} compared the effective radius $\reff$ of CARMA with \cite{ackerman2001cloud} and found that for CARMA, $\reff$ attain its maximum near to the cloud base, whereas the \cite{ackerman2001cloud} model produces a roughly constant $\reff$ profile with depth.

To study how parameters $\alpha$ and $\beta$ of the variable $\fsed$ function \eqref{eq:fsed} influence the cloud mass mixing ratio and the effective particle radius, we considered four representative cases in Figure \ref{fig:effect_of_params}.
In particular, we plot the profiles for $\alpha=0.5,5$ and $\beta=0.4,10$, noting that the $\beta=10$ cases are effectively constant $\fsed=\alpha$.
We notice that the large $\beta=10$ profiles produce effective radii profiles similar to that reported by \cite{gao2018sedimentation}, namely those that do not vary greatly with depth.
The larger $\alpha=5$ profile results in larger effective radii than $\alpha=0.5$, which we expect from \cite{ackerman2001cloud} and \cite{gao2018sedimentation}.
For $\beta=0.4$, the $\reff$ profiles indicate that the maximum effective radius is obtained closer to the top of the atmosphere than the bottom of the cloud deck, which is an expected result.
Due to the understanding that larger $\fsed$ values produce larger particle sizes, it is reasonable that an $\fsed$ profile that increases with altitude would create an $\reff$ profile that also increases with altitude. 
However, this conflicts with the behaviour of CARMA reported by \cite{gao2018sedimentation}.

More work is needed to explore the impact of variable $\fsed$ on the particle size distributions and will be the subject of a future publication. In this proof-of-concept study we focus only on fitting the cloud mass mixing ratio of $\virga$ to that of CARMA.

\subsection{Effect on spectra}
\begin{figure*}[b!]
	\centering
	\gridline{\fig{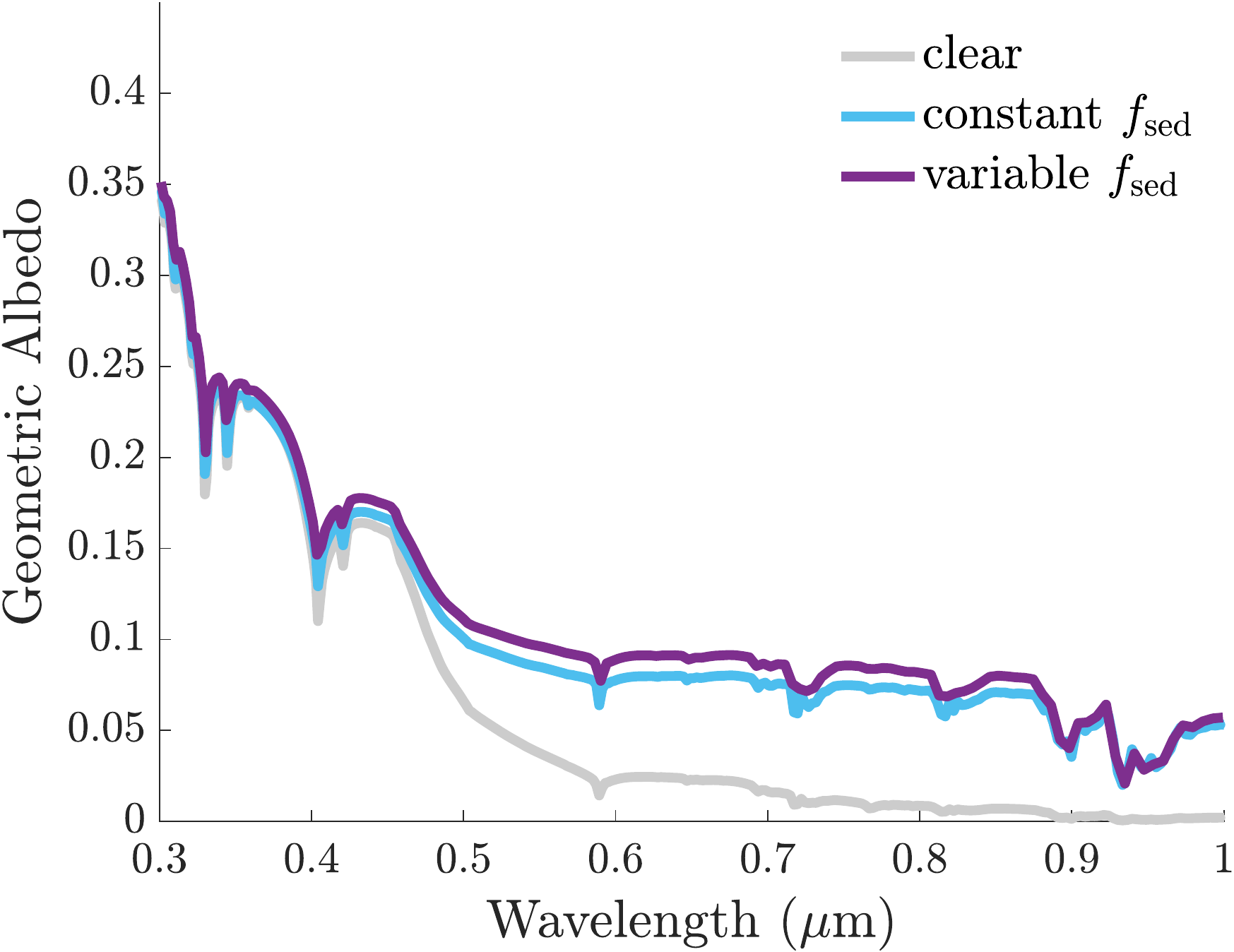}{0.45\textwidth}
					{(a) Heterogeneous nucleation fit with CCN flux = 100 cm$^{-2}$s$^{-1}$ (parameters outlined in Figure \ref{fig:fig7_qc_comparison}(a))}
					\fig{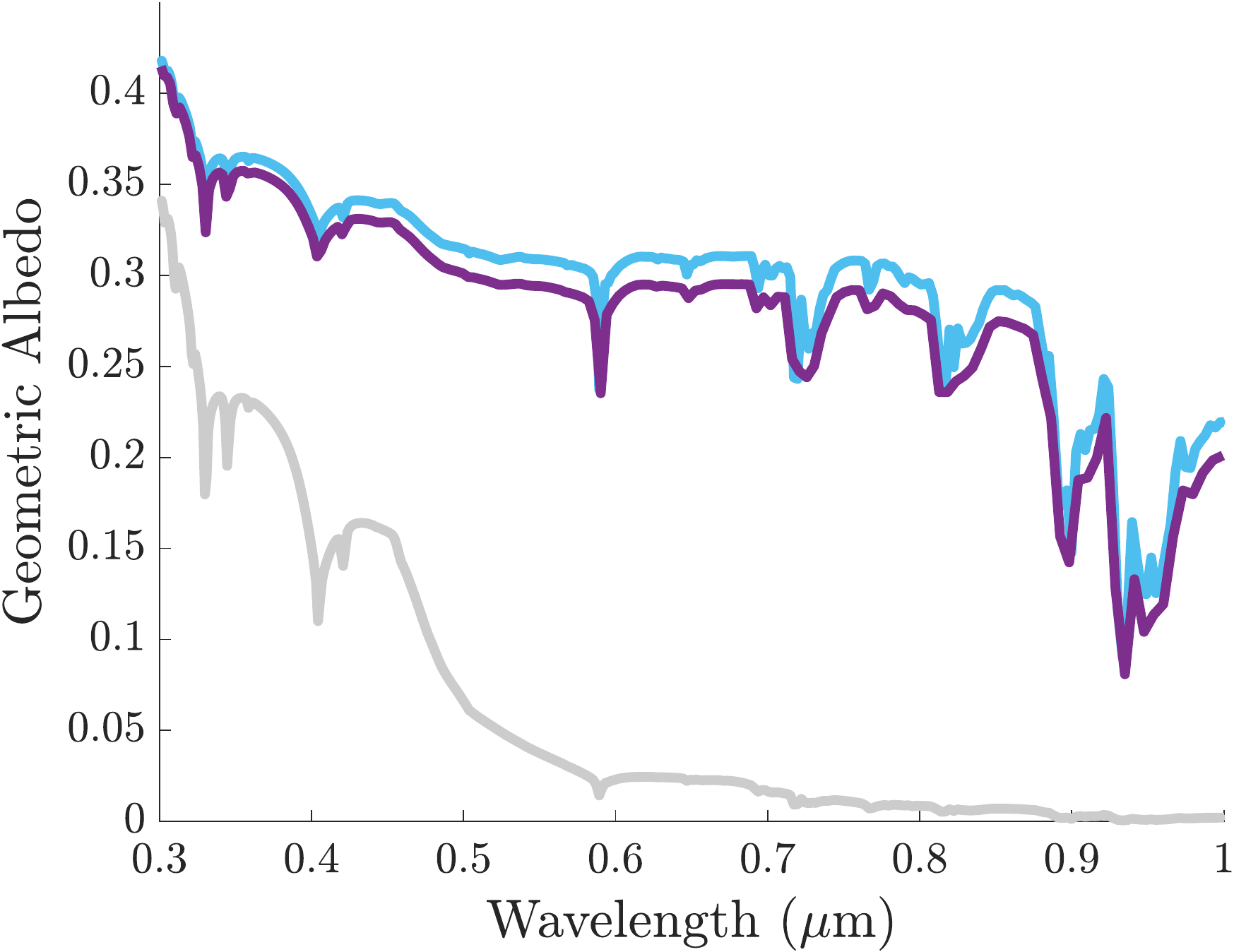}{0.45\textwidth}
					{(b) Heterogeneous nucleation fit with CCN flux = 1000 cm$^{-2}$s$^{-1}$ (parameters outlined in Figure \ref{fig:fig7_qc_comparison}(b))}
          }
	\gridline{\fig{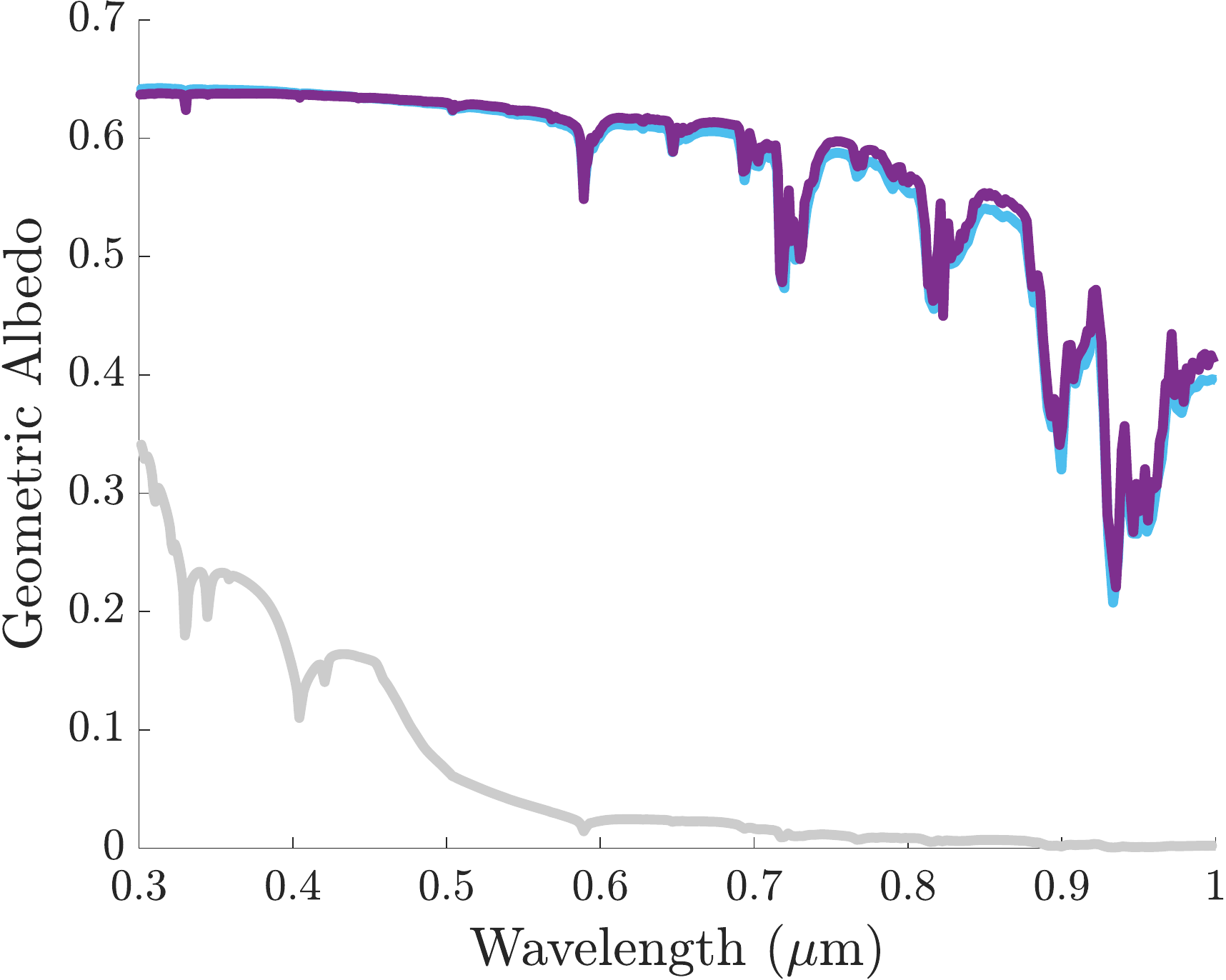}{0.45\textwidth}
				{(c) Homogeneous nucleation fit with surface energy = $\sigma_0^\text{KCl}/4.$ (parameters outlined in Figure \ref{fig:fig4_qc_comparison}(a))}
				\fig{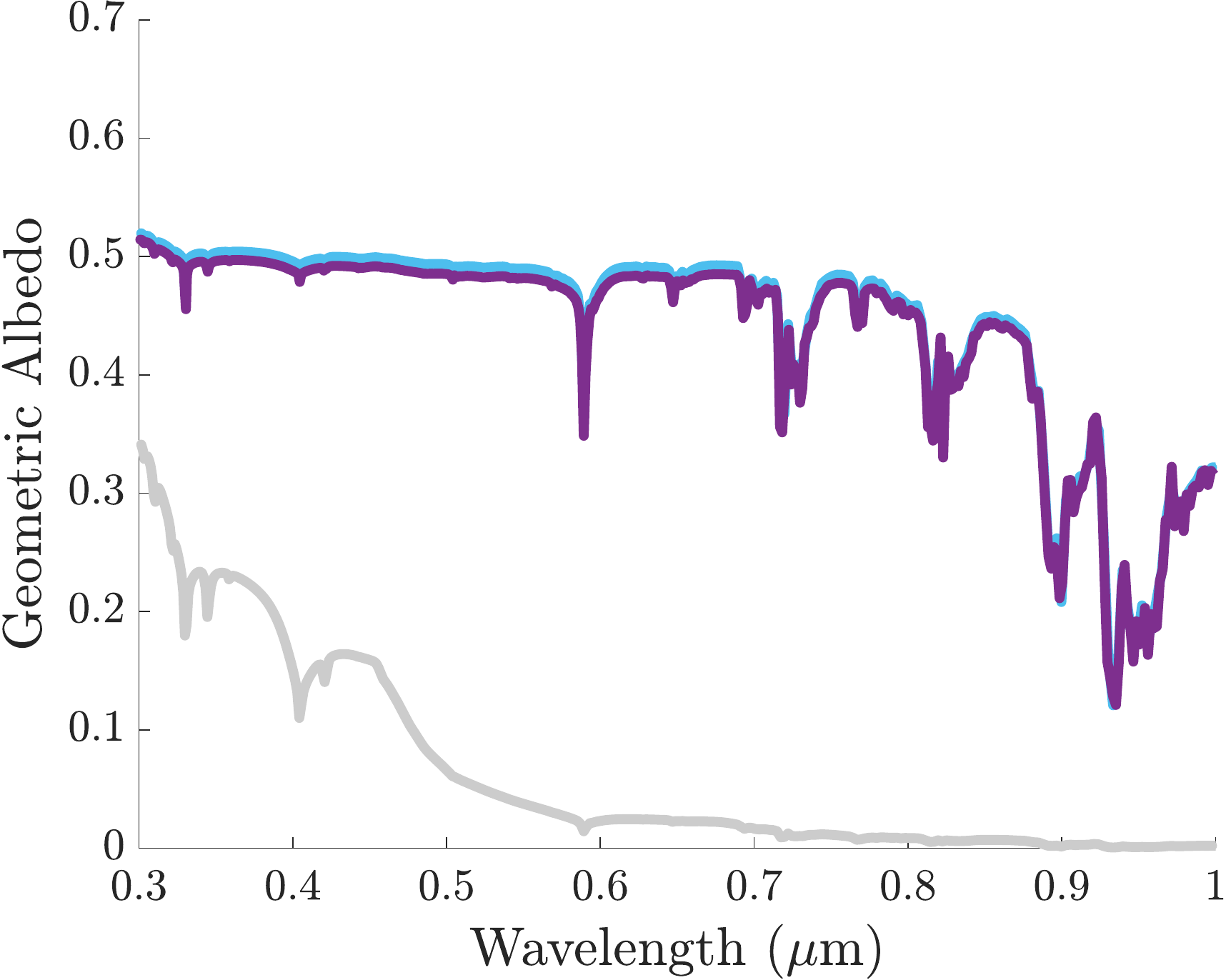}{0.45\textwidth}
				{(d) Homogeneous nucleation fit with surface energy = $2\sigma_0^\text{KCl}$ (parameters outlined in Figure \ref{fig:fig4_qc_comparison}(b))}
          }
	\caption{Reflected light spectra produced by constant and variable $\fsed$ profiles in $\virga$ for a H$_2$/He-dominated atmosphere with solar metallicity, $T_\mathrm{eff}=400K$ and $\log g=5.25$, for KCl clouds with $\kzz=10^8$~cm$^2$~s$^{-1}$ for heterogeneous nucleation and $\kzz=10^7$~cm$^2$~s$^{-1}$ for homogeneous nucleation. Atmospheres originate from \cite{marley2018sonora}.}
	\label{fig:reflected_spectra_het_hom}
\end{figure*}

\begin{figure*}[b!]
	\centering
	\gridline{\fig{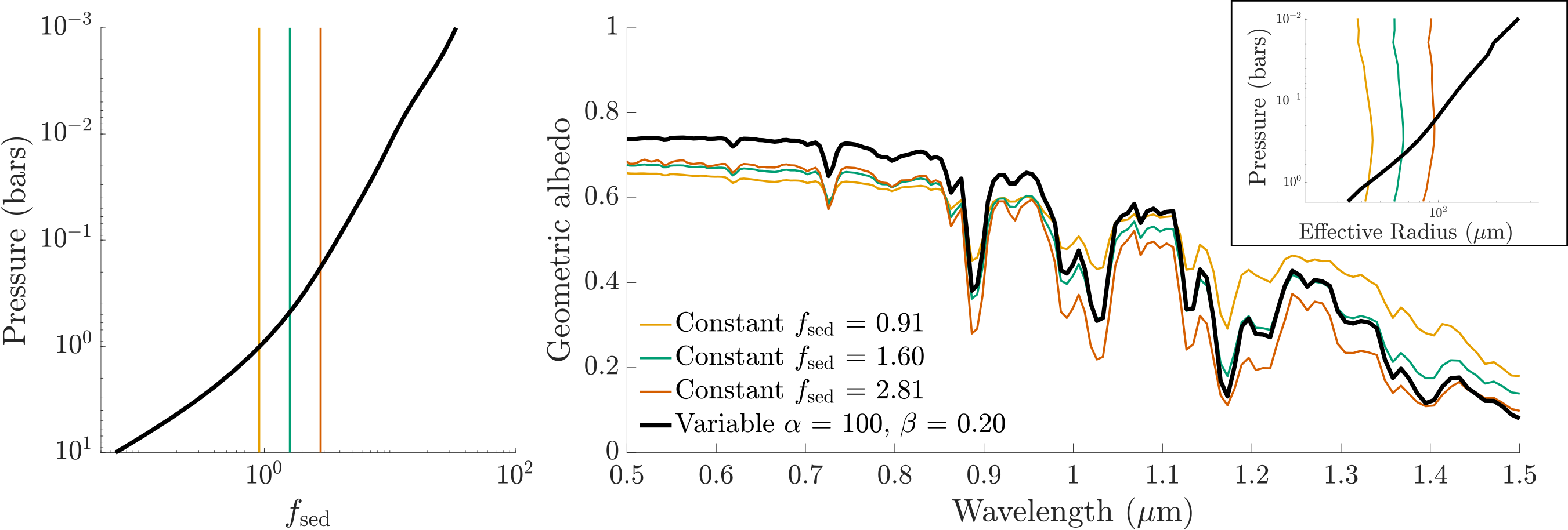}{0.9\textwidth}
					{(a) Reflected light spectra.}
          }
    \gridline{\fig{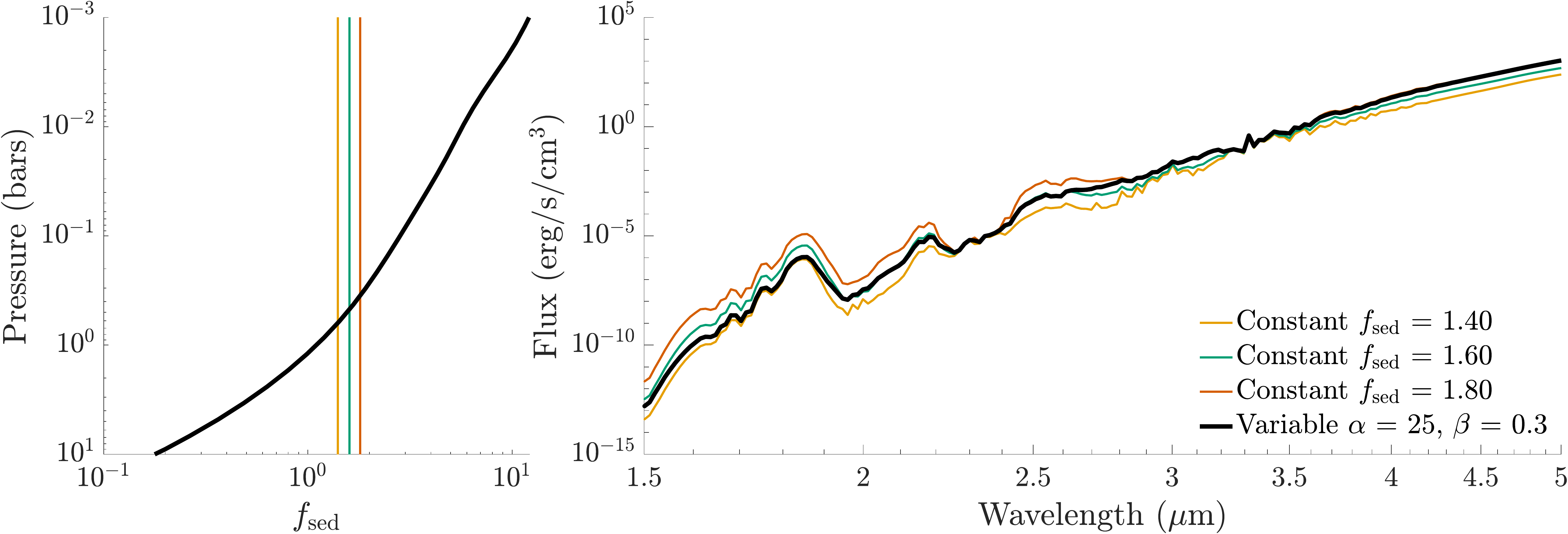}{0.9\textwidth}
					{(b) Thermal spectra.}
          }
	\caption{Reflected and thermal spectra for a Jupiter-like planet with H$_2$O and NH$_3$ clouds. The left-hand panels depict the variation of $\fsed$ with pressure, whereas the right-hand panels illustrate the resulting spectra for each choice of $\fsed$.}
	\label{fig:jupiter_spectra}
\end{figure*}

The complexity added to \texttt{virga} can be used to more closely match to observations. Therefore, in this final section we demonstrate how our new methodology is expected to influence spectroscopy. We first focus on the cases explored above. Then, more generally, we demonstrate how variable $\fsed$ will impact spectroscopy. Specifically, we focus on reflected light and thermal emission since they are more heavily dependent on the distinct scattering nature of the cloud deck. Ultimately our framework allows for the computation of transmission, as well. We use the \texttt{picaso} \citep{picaso} radiative transfer suite, with the v2 opacity database \citep{picaso_opacity} that are thoroughly described in \citep{marley2018sonora}. For atmospheric chemical abundances needed for spectroscopy we use the abundances of \citet{marley2018sonora} which are based on the methodology of \citet{visscher2006atmospheric}. 

In Figure \ref{fig:reflected_spectra_het_hom} we plot the reflected light spectra for the cases discussed in Section \ref{sec:carma_comparison}: H$_2$/He-dominated atmosphere with solar metallicity, $T_\mathrm{eff}=400K$ and $\log g=5.25$, with KCl clouds. 
While this particular underlying model does not represent a realistic irradiated object, the reflected light calculation illustrates the sensitivity of observable quantities to the details of the cloud structure.
To illustrate the differences, we plot the reflected light spectra obtained using constant and variable $\fsed$ profiles obtained through best-fit with the CARMA cloud mass mixing ratios for heterogeneous and homogeneous nucleation (Figures \ref{fig:reflected_spectra_het_hom}(a)\&\ref{fig:reflected_spectra_het_hom}(b) and \ref{fig:reflected_spectra_het_hom}(c)\&\ref{fig:reflected_spectra_het_hom}(d) respectively).

As expected, the homogeneous nucleation case produce near identical spectra. Very minor differences in the homogeneous case are due to the computed single scattering albedo profiles, which differ by a maximum of 0.1 -- enough to impact the scattering properties of the observed spectrum.  For heterogeneous nucleation the spectra produced using variable and constant $\fsed$ profiles are more disparate but relatively minor (A$_g<$0.025) considering the observational capability of future flagship missions \citep{Feng2018AJ}. In this case, the minor impact to the spectrum is a result of the low total optical depth for the heterogeneous nucleation case (P($\tau\sim$=0.1)=1 bar). Many exoplanets and brown dwarfs have been hypothesized to have optically thick cloud decks at much lower pressures \citep[e.g.,][]{webber2015effect}. Therefore, we move towards showing the general effect of variable $\fsed$ on a benchmark system computed from the self-consistent models in \cite{picaso}, it is a $25\,\rm m\,s^{-2}$ Jupiter-like planet with H$_2$O and NH$_3$ clouds, 5~AU from a Sun-like star with 3$\times$M/H.
This atmosphere scenario was included as one of the original benchmark systems used for comparison to other codes\footnote{See \texttt{picaso.justdoit.jupiter\_pt}}, therefore, it serves as a an ideal candidate to present sensitivity tests.

We compare the reflected (Figure \ref{fig:jupiter_spectra}(a)) and thermal (Figure \ref{fig:jupiter_spectra}(b)) spectra 
for a number of constant $\fsed$ values and a comparable variable $\fsed$ function.
On the left-hand panel of each figure, we plot $\fsed$ with pressure to illustrate how the variable $\fsed$ function varies compared to the constant values chosen.
The spectra corresponding to each choice of $\fsed$ is plotted on the right-hand panel, along with an effective radius inset for the reflected light case. 
For the variable $\fsed$ profiles given by \eqref{eq:fsed}, we use $\epsilon=10^{-2}$, $P(z=z_T)=10^{-4}$ bar whereas $\alpha$ and $\beta$ are indicated in the spectral plots in Figure \ref{fig:jupiter_spectra}.

For thermal emission, the constant $\fsed$ spectra monotonically decrease in flux as $\fsed$ decrease. This is because the decreasing $\fsed$ pushes the $\tau=1$ pressure-level surface toward lower pressures. Therefore, the lowest $\fsed$ spectrum, which produces a more vertically extended cloud deck, consistently probes cooler temperatures resulting in lower fluxes at all wavelengths. In reflected light, the picture is complicated by confounding Rayleigh scattering short-ward of 1$\mu$m. A pure-Rayleigh scattering atmosphere is expected to approach 0.75 \citep{horak1950diffuse}. Therefore short-ward of 1$\mu$m where Rayleigh scattering is dominant, the spectra are expected to brighten with increasing $\fsed$ tending towards that of a clear atmosphere. Long-ward of 1$\mu$m where the interplay is instead between molecular absorption and scattering by cloud opacity, a pure-absorbing atmosphere would approach zero reflectively. Therefore the spectra are expected to darken with increasing $\fsed$.  

From Figure \ref{fig:jupiter_spectra}(a), we observe that the variable $\fsed$ spectra tells a more complex story because now the particle radii change with altitude which in turn alters the scattering properties with altitude. In reflected light, the variable-$\fsed$ profile is generally brighter than the others for wavelengths less than 1~$\mu$m, where Rayleigh scattering is dominant. The exception to this is within the optically thick H$_2$O band (0.9$\mu$m), where the variable $\fsed$ spectrum becomes as dark as the constant $\fsed$=0.91 case. As wavelength increases (1-1.5$\mu$m) the variable-case further interlaces with the constant $\fsed$ cases. In the regions that are moderately dominated by molecular absorption (i.e. where geometric albedo a$_g$~$\sim$~0.2-0.55), the variable-$\fsed$ spectrum follows the constant $\fsed$ profile of 1.6. In the darker regions that are heavily dominated by molecular absorption (where a$_g<0.2$), the variable-case spectrum follows the higher $\fsed=2.81$. 

Overall this is an intuitive, yet powerful result. Darker regions of the reflected light spectrum probe lower pressures, as the $\tau=1$ pressure-level of the molecular opacity moves to lower pressures. Therefore, with a variable $\fsed$ profile each molecular feature in the spectrum is perturbed according to the cloud model of the corresponding pressure-level it is sensitive to. Such a spectrum could not be obtained by a constant $\fsed$, therefore this greater flexibility expands the range of spectra that can be produced by $\virga$. 

Similar behaviour is evident for the thermal spectra in Figure \ref{fig:jupiter_spectra}(b). The variable $\fsed$ profile approaches a higher constant-$\fsed$ profile in the regions relatively void of molecular opacity, i.e. the 3-5$\mu$m region that resembles a blackbody. On the other hand, in the optically-thick band centers that probe lower pressures, the variable case spectrum approaches the lower constant $\fsed$ profiles.

Given the effect seen on spectra, it is clear that the variable $\fsed$ methodology will be  especially critical for fitting observations that span a wide wavelength range. Overall, an obvious diagnostic of an observation that would be suited for this variable $\fsed$ methodology is a spectra that shows clear influence from clouds but whose molecular features cannot be accurately fit across a wide wavelength region.  For example, in our reflected light case in Figure \ref{fig:jupiter_spectra}(a), the water band at 0.9$\mu$m would need to be fit by a constant $\fsed=1.6$ cloud model whereas the 1.4$\mu$m-band would need to be fit by a constant $\fsed=2.81$ cloud model. Instead, our methodology would allow the entire spectrum to be fit uniformly across wavelength.

\section{Conclusions}
\label{sec:conclusions}
The parametric cloud sedimentation model $\virga$ \cite{virga}, originating from the work of \cite{ackerman2001cloud}, is a powerful tool for predicting the vertical extent of condensate clouds in exoplanet and brown dwarf atmospheres.
Based upon a balance between turbulent mixing and sedimentation, the model neglects all microphysical processes except for condensation, making it extremely computationally efficient yet physically insightful.
Its numerical efficacy is contingent upon the availability of an analytical solution to the diffusion-sedimentation equation \eqref{eq:diff-sed-eqn}, thus $\virga$ followed the \cite{ackerman2001cloud} methodology by enforcing a constant sedimentation efficiency $\fsed$.
However, analytical capability is not dependent upon a constant $\fsed$, we simply require that $\fsed$ has an explicit anti-derivative. 
We thereby extended $\virga$ by allowing $\fsed$ to vary exponentially with altitude throughout the atmosphere.
This modification introduces a plethora of cloud profiles, and consequently atmospheric spectra, that the constant $\fsed$ implementation of $\virga$ is unable to produce.

As a preliminary investigation into the impact of this extension, we revisited the analysis of \cite{gao2018sedimentation}, who compared the cloud mass mixing ratios produced by \cite{ackerman2001cloud} with that of CARMA, a microphysically intricate cloud model, to elucidate the atmospheric properties upon which $\fsed$ was most dependent.
In this original work, the cloud profiles produced by each models were rarely in agreement.
We reproduced two of the investigations of \cite{gao2018sedimentation}, namely cases involving heterogeneous and homogeneous nucleation, to explore whether our new variable $\fsed$ implementation might improve agreement between the two models.

From the comparisons conducted in Section \ref{sec:carma_comparison}, it is clear that an altitude-dependent expression for sedimentation efficiency $\fsed$ introduces greater diversity in cloud mass mixing ratio profiles.
In particular, it is possible that we can better replicate the cloud mass mixing ratio produced by heterogeneous nucleation in CARMA, a microphysical process that is not directly included in \texttt{virga}.
We suggest a possible reason for this improved comparison is the small number of nucleation centers at the top of the cloud in CARMA due to the relatively low CCN flux. 
This results in low cloud particle densities, however these particles can grow to relatively large sizes via condensation, since all of the available vapour will be shared among only a few particles.
Consequently, this altitude-dependent CCN distribution will be favoured by an altitude-dependent $\fsed$, namely one that increases with altitude.

On the other hand, there appears to be little benefit of a variable $\fsed$ function when improving the fit of \texttt{virga} to CARMA in the case of homogeneous nucleation.
This is likely due to the fact that homogeneous nucleation is independent of CCNs but dependent on the nucleation energy barrier, where the latter is a function of material properties and supersaturation.
The low nucleation energy barrier of KCl allows nucleation to proceed quickly, therefore the cloud distribution is primarily defined by condensational growth and particle transport. 
A constant $\fsed$ captures such effects adequately.
However, it is possible that a different variable $\fsed$ formulation might improve upon the constant $\fsed$ fit.

Finally, we demonstrated how a variable $\fsed$ function affects the reflected light and thermal emission spectroscopy of a Jupiter-like planet with H$_2$O and NH$_3$ clouds.
We compared the spectra for a number of constant $\fsed$ values and a variable $\fsed$ function to illustrate the greater diversity offered by the latter.
We observed that with a variable $\fsed$ profile each molecular feature in the spectrum is perturbed according to the constant $\fsed$ model of the corresponding pressure-level it is sensitive to. 
Similar behaviour is observed for the thermal spectra.

We have considered only one example of a variable $\fsed$ function derived from the atmospheric density dependence of the sedimentation velocity and eddy-diffusion coefficient $\kzz$.
However, we are far from restricted to this choice;
\texttt{virga} possesses the capability to handle any variable $\fsed$ function so long as it has an explicit anti-derivative expression.
Alternative choices for $\fsed$ will impact not only the cloud mass mixing ratio but also the particle size distribution, therefore there is significant opportunity to vary this function and explore the impact on the cloud properties.
The considerable flexibility allowed by this $\fsed$ function paves the way towards an abundance of different cloud profiles, and consequently different spectral behaviours, that are otherwise unattainable by the constant $\fsed$ profiles we have been restricted to thus far.

\begin{acknowledgements}
C.R.’s research was supported by an appointment to the NASA Postdoctoral Program at the NASA Ames Research Center, administered by Universities Space Research Association under contract with NASA. N.B. acknowledges support from the NASA Astrophysics Division and M.M. acknowledges support from the NASA Planetary Science Division.
\end{acknowledgements}

\software{numba \citep{numba}, pandas \citep{mckinney2010data}, bokeh \citep{bokeh}, NumPy \citep{walt2011numpy}, IPython \citep{perez2007ipython}, Jupyter, \citep{kluyver2016jupyter}, Virga \citep{virga}, PICASO \citep{picaso}, MATLAB \citep{MATLAB:2010}}

\clearpage
\bibliographystyle{aasjournal}
\bibliography{main}

\end{document}